\documentclass{aa}

\usepackage{graphicx}
\usepackage{natbib}
\usepackage{txfonts}

\bibpunct{(}{)}{;}{a}{}{,}

\begin{document}

\title{Extended warm gas in Orion KL\\as probed by methyl cyanide\thanks{Based on observations carried out with the IRAM 30m telescope. IRAM is supported by INSU/CNRS (France), MPG (Germany), and IGN (Spain).}}
\titlerunning{Extended warm gas in Orion KL probed by methyl cyanide}
\authorrunning{Bell et al.}

\author{
 T.~A.~Bell\inst{1} \and
 J.~Cernicharo\inst{1} \and
 S.~Viti\inst{2} \and
 N.~Marcelino\inst{3} \and
 Aina Palau\inst{4} \and
 G.~B.~Esplugues\inst{1} \and
 B.~Tercero\inst{1}
}

\institute{
 Centro de Astrobiolog\'ia (CSIC-INTA), Torrej\'on de Ardoz, 28850 Madrid, Spain\\
 \email{tab@cab.inta-csic.es}
\and
 Department of Physics and Astronomy, University College London, Gower Street, London, WC1E 6BT, UK
\and
 National Radio Astronomy Observatory, 520 Edgemont Road, Charlottesville, VA 22903, USA
\and
 Institut de Ci\`encies de l'Espai (CSIC-IEEC), Campus UAB, Facultat de Ci\`encies, Torre C-5 parell 2, E-08193 Bellaterra, Spain
}

\date{Received 2013 / Accepted 2014}

\abstract{
In order to study the temperature distribution of the extended gas within the Orion Kleinmann-Low nebula, we have mapped the emission by methyl cyanide (CH$_3$CN) in its $J$=$6_K$--$5_K$, $J$=$12_K$--$11_K$, $J$=$13_K$--$12_K$, and $J$=$14_K$--$13_K$ transitions at an average angular resolution of $\sim$10 arcsec (22 arcsec for the $6_K$--$5_K$ lines), as part of a new 2D line survey of this region using the IRAM 30m telescope. These fully sampled maps show extended emission from warm gas to the northeast of IRc2 and the distinct kinematic signatures of the hot core and compact ridge source components. We have constructed population diagrams for the four sets of $K$-ladder emission lines at each position in the maps and have derived rotational excitation temperatures and total beam-averaged column densities from the fitted slopes. In addition, we have fitted LVG model spectra to the observations to determine best-fit physical parameters at each map position, yielding the distribution of kinetic temperatures across the region. The resulting temperature maps reveal a region of hot ($T > 350$~K) material surrounding the northeastern edge of the hot core, whereas the column density distribution is more uniform and peaks near the position of IRc2. We attribute this region of hot gas to shock heating caused by the impact of outflowing material from active star formation in the region, as indicated by the presence of broad CH$_3$CN lines. This scenario is consistent with predictions from C-shock chemical models that suggest that gas-phase methyl cyanide survives in the post-shock gas and can be somewhat enhanced due to sputtering of grain mantles in the passing shock front.
}


\maketitle


\section{Introduction}\label{Introduction}

The Orion Kleinmann-Low (KL) nebula is the closest \citep[414 pc,][]{Menten2007} and most studied high-mass star-forming region in the Galaxy. The prevailing chemistry in this source is particularly complex as a result of the interactions between newly formed protostars, their outflows and their environment. The evaporation of dust mantles and the presence of high gas temperatures produce a wide variety of molecules in the gas phase that are responsible for a spectacularly rich and intense line spectrum \citep{Blake1987, Charnley1997}. The chemical complexity of Orion KL has been demonstrated by an extensive number of line surveys, performed over a variety of frequency ranges from 72 GHz up to 1.9 THz \citep[e.g.,][]{Blake1986, Schilke1997, Bergin2010a}. Finally, millimetre and submillimetre aperture synthesis studies have provided the spatial location and extent of many molecular species \citep[e.g.,][]{Blake1996, Beuther2005, Plambeck2009, Zapata2009}, further indicating the presence of distinct physical/kinematic components that show clear chemical differentiation, with NH- and OH-bearing molecules peaking in different sources (the hot core and compact ridge, respectively).

Near- and mid-IR subarcsecond resolution imaging and (sub)millimetre interferometric observations have identified the main sources of luminosity, heating, and dynamics in the region \citep[e.g.,][]{Dougados1993, Gezari1998, Shuping2004, DeBuizer2012}. However, the nature of these objects has been the subject of much debate, as has the question of which source(s) are most responsible for the high luminosity produced by the nebula \citep[$\sim$10$^5$ $L_{\sun}$;][]{Menten1995}. The main sources of activity seem to be radio sources \textit{I} and \textit{n}, located just to the south and southwest of IRc2. The Becklin-Neugebauer (BN) object 9$\arcsec$ northwest of the hot core is believed to be internally heated by an embedded massive protostar (or protostars). Proper motion studies have shown that all three sources, \textit{I}, \textit{n}, and BN, are receding from a common origin. This has led to the hypothesis that they were expelled from a massive young stellar system that disintegrated $\sim$500 years ago \citep{Bally2005, Zapata2009, Niederhofer2012, Nissen2012}, and that this explosive event may be the source of heating for the region in and around the hot core \citep{Zapata2009, Zapata2011, Moeckel2012}. The spectacular molecular ``fingers'' that trace the high-velocity outflow emanating from KL in a northwest-southeast direction have also been attributed to this explosive event.

Interferometric observations probing the central region of the KL nebula at arcsecond resolution have also revealed another distinct kinematic component, particularly evident in shock tracers such as SiO and SO, that extends in the northeast-southwest direction and has been identified as a low-velocity bipolar outflow originating at source \textit{I} \citep[e.g.,][]{Plambeck1982, Greenhill1998, Niederhofer2012}. This, along with the spectral shape of its continuum emission, has therefore lead various authors to argue that source \textit{I} is likely to be an embedded protostar driving this outflow. Zapata et al. (2009, 2011) have proposed instead that the low-velocity outflow is linked to the explosive event and that the region surrounding IRc2 and the hot core is being externally heated by material ejected from the explosion impacting the stationary gas. The nature of the Orion hot core, BN object, and sources \textit{I} and \textit{n}, and the underlying cause of the dynamics and energetics in the KL nebula are thus matters of continued debate.

Reliable measurements of the temperature distribution across the extended gas in the KL nebula can help to better understand the energetics in the region, and therefore the likely driving sources. Methyl cyanide, CH$_3$CN, is an excellent probe of the gas temperature in warm dense environments like Orion KL \citep{Boucher1980}. Its molecular symmetry gives rise to many transitions closely spaced in frequency that span a large range of energies (up to $E_\mathrm{up} \! > \! 1000$ K). These lines can therefore be observed simultaneously at one frequency setting, thus avoiding flux calibration uncertainties, and are excited by collisions (so-called $K$-ladder transitions) so that their level populations are dictated by the gas temperature. Nitrogen-bearing complex organic molecules such as methyl cyanide are generally assumed to form on grain mantles in the cold dense phase of the interstellar medium, before being released into the gas phase by evaporation in the hot regions surrounding massive protostars. Their spatial distribution and abundances can therefore provide important constraints on chemical models of these sources.

The methyl cyanide emission in Orion KL has been studied extensively, both with single-dish and interferometric observations. Early observations of the millimetre lines of CH$_3$CN \citep{Loren1981, Loren1984, Sutton1986} revealed the presence of warm gas ($\sim$275~K) in the hot core and weaker extended emission from cooler gas ($\sim$100~K) in the quiescent ridge, with derived fractional abundances of 10$^{-11}$ to 10$^{-10}$. More recent interferometric studies \citep{Wilner1994, Beuther2006, Wang2010} have uncovered the complex and clumpy structure within $\sim$10$\arcsec$ of IRc2, yielding kinetic temperatures above 250~K (and even up to 600~K, based on submillimetre transitions) for the hot core and lower values of $\sim$150--250~K toward the compact ridge. These arcsecond resolution observations suggest much higher CH$_3$CN abundances of 10$^{-8}$ to 10$^{-7}$ in the hot core and compact ridge.

We have obtained fully sampled maps of four sets of methyl cyanide $K$-ladder emission lines across a large region centred on Orion IRc2. These maps cover a larger area with improved sensitivity and (single-dish) spatial resolution than those of previous studies, providing a homogeneous dataset that is well-suited to studying the extended emission in the region surrounding the KL nebula. Furthermore, while these data cannot compete with the angular resolution afforded by interferometric techniques, they do not suffer from the problem of missing flux common to many interferometric observations, which tend to emphasise the clumpiness of the gas by filtering out the smooth extended emission. Recovering this extended emission is crucial when studying the structure on larger scales, which is the focus of this paper. Based on these data, we have analysed the kinematics of this extended warm gas and derived maps of temperature and column density across the region. We describe the observations that were obtained and the subsequent reduction of those data in Sect.~\ref{Observations}. The maps are presented in Sect.~\ref{Results:Emission-Maps} and the general kinematic properties of the emitting gas are discussed. We describe the analysis of these maps and the derivation of column densities and rotational temperatures by use of the population diagram method, followed by LVG model fitting in Sect.~\ref{Analysis}. The resulting temperature distributions are presented in Sect.~\ref{Results:Temperature-Maps} and their implications for the source structure are discussed in Sect.~\ref{Discussion}. Our findings are summarised in Sect.~\ref{Conclusions}.


\section{Observations and data reduction}\label{Observations}

\begin{table}
 \caption{Spectroscopic properties of the observed methyl cyanide lines.}
 \label{Table:Line-Properties}
 \centering
 \begin{tabular}{c@{\ \ }cccrr}
 \hline\hline
 \multicolumn{2}{c}{Transition} & Frequency & $A_{i\!j}$ &
 \multicolumn{1}{c}{$E_\mathrm{up}$} & \multicolumn{1}{c}{$g_\mathrm{up}$} \\
 $J'$$\to$$J''$ & $K$ & (GHz) & (s$^{-1}$) & \multicolumn{1}{c}{(K)} & \\
 \hline
  6--5  &  0 & 110.383500 & 1.11$\times$10$^{-4}$ &  18.5 &  26 \\
  6--5  &  1 & 110.381372 & 1.08$\times$10$^{-4}$ &  25.7 &  26 \\
  6--5  &  2 & 110.374989 & 9.88$\times$10$^{-5}$ &  47.1 &  26 \\
  6--5  &  3 & 110.364354 & 8.33$\times$10$^{-5}$ &  82.8 &  52 \\
  6--5  &  4 & 110.349470 & 6.17$\times$10$^{-5}$ & 132.8 &  26 \\
  6--5  &  5 & 110.330345 & 3.39$\times$10$^{-5}$ & 197.1 &  26 \\
 \hline
 12--11 &  0 & 220.747261 & 9.24$\times$10$^{-4}$ &  68.9 &  50 \\
 12--11 &  1 & 220.743011 & 9.18$\times$10$^{-4}$ &  76.0 &  50 \\
 12--11 &  2 & 220.730261 & 8.99$\times$10$^{-4}$ &  97.4 &  50 \\
 12--11 &  3 & 220.709016 & 8.66$\times$10$^{-4}$ & 133.2 & 100 \\
 12--11 &  4 & 220.679287 & 8.21$\times$10$^{-4}$ & 183.1 &  50 \\
 12--11 &  5 & 220.641084 & 7.63$\times$10$^{-4}$ & 247.4 &  50 \\
 12--11 &  6 & 220.594423 & 6.92$\times$10$^{-4}$ & 325.9 & 100 \\
 12--11 &  7 & 220.539323 & 6.08$\times$10$^{-4}$ & 418.6 &  50 \\
 12--11 &  8 & 220.475807 & 5.12$\times$10$^{-4}$ & 525.6 &  50 \\
 12--11 &  9 & 220.403900 & 4.03$\times$10$^{-4}$ & 646.7 & 100 \\
 12--11 & 10 & 220.323631 & 2.81$\times$10$^{-4}$ & 782.0 &  50 \\
 \hline
 13--12 &  0 & 239.137916 & 1.18$\times$10$^{-3}$ &  80.3 &  54 \\
 13--12 &  1 & 239.133313 & 1.17$\times$10$^{-3}$ &  87.5 &  54 \\
 13--12 &  2 & 239.119504 & 1.15$\times$10$^{-3}$ & 108.9 &  54 \\
 13--12 &  3 & 239.096497 & 1.12$\times$10$^{-3}$ & 144.6 & 108 \\
 13--12 &  4 & 239.064299 & 1.07$\times$10$^{-3}$ & 194.6 &  54 \\
 13--12 &  5 & 239.022924 & 1.00$\times$10$^{-3}$ & 258.9 &  54 \\
 13--12 &  6 & 238.972389 & 9.26$\times$10$^{-4}$ & 337.4 & 108 \\
 13--12 &  7 & 238.912715 & 8.35$\times$10$^{-4}$ & 430.1 &  54 \\
 13--12 &  8 & 238.843926 & 7.30$\times$10$^{-4}$ & 537.0 &  54 \\
 13--12 &  9 & 238.766049 & 6.11$\times$10$^{-4}$ & 658.2 & 108 \\
 13--12 & 10 & 238.679115 & 4.79$\times$10$^{-4}$ & 793.4 &  54 \\
 \hline
 14--13 &  0 & 257.527383 & 1.48$\times$10$^{-3}$ &  92.7 &  58 \\
 14--13 &  1 & 257.522427 & 1.47$\times$10$^{-3}$ &  99.8 &  58 \\
 14--13 &  2 & 257.507561 & 1.45$\times$10$^{-3}$ & 121.3 &  58 \\
 14--13 &  3 & 257.482792 & 1.41$\times$10$^{-3}$ & 157.0 & 116 \\
 14--13 &  4 & 257.448128 & 1.35$\times$10$^{-3}$ & 207.0 &  58 \\
 14--13 &  5 & 257.403584 & 1.29$\times$10$^{-3}$ & 271.2 &  58 \\
 14--13 &  6 & 257.349179 & 1.20$\times$10$^{-3}$ & 349.7 & 116 \\
 14--13 &  7 & 257.284935 & 1.10$\times$10$^{-3}$ & 442.4 &  58 \\
 14--13 &  8 & 257.210877 & 9.90$\times$10$^{-4}$ & 549.4 &  58 \\
 14--13 &  9 & 257.127035 & 8.62$\times$10$^{-4}$ & 670.5 & 116 \\
 14--13 & 10 & 257.033444 & 7.19$\times$10$^{-4}$ & 805.8 &  58 \\
 \hline
 \end{tabular}
 \tablefoot{Spectroscopic data taken from the Cologne Database for Molecular Spectroscopy \citep[CDMS;][]{Muller2005}. $A_{i\!j}$ is the Einstein coefficient for spontaneous emission, $E_\mathrm{up}$ is the upper state energy, and $g_\mathrm{up}$ is the upper state degeneracy.}
\end{table}

\begin{figure*}
 \centering
 \includegraphics[width=17cm]{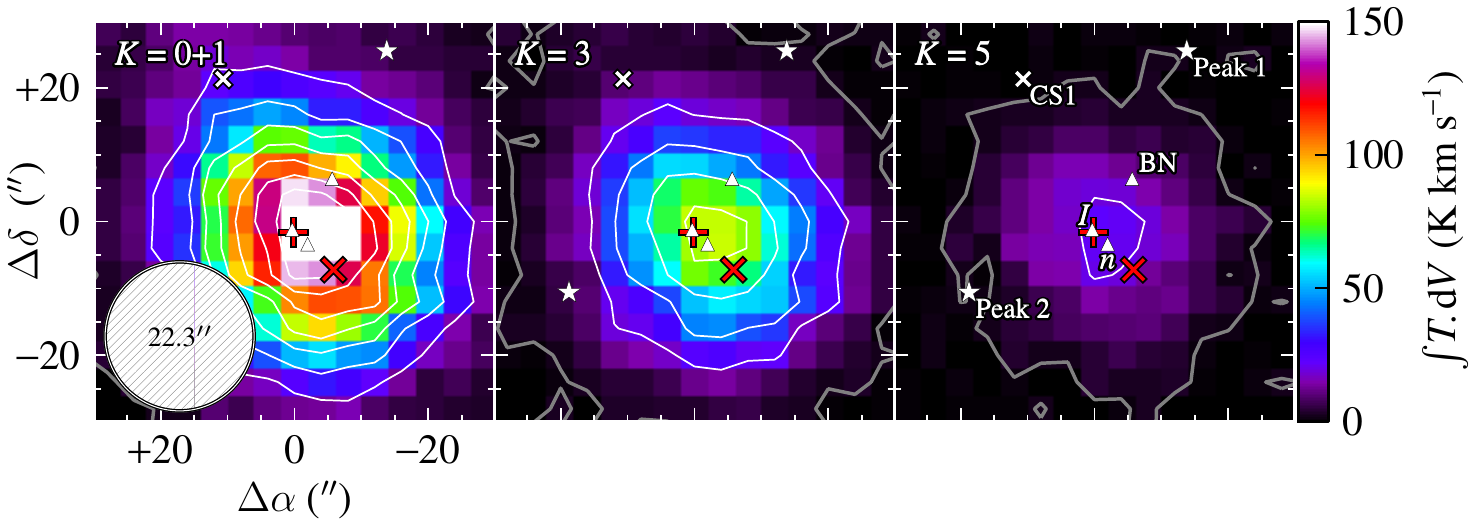}
 \caption{Integrated intensity maps of methyl cyanide $J$=$6_K$--$5_K$ emission across the Orion KL region. Contours are plotted at the 3-sigma level ($3\sigma = 3 \!\times\! T_\mathrm{rms} \, \delta V \sqrt{N}_\mathrm{channels} = 2.1$~K\,km\,s$^{-1}$; grey lines) and from 20 to 140 in 20~K\,km\,s$^{-1}$ intervals (white lines). The positions of the hot core and compact ridge are indicated by a red plus sign and a diagonal cross, respectively. The locations of radio continuum sources BN, \textit{I}, and \textit{n} are also marked on each panel by white triangles. Peaks 1 and 2 of the H$_2$ emission are indicated by white stars and the condensation CS1 is marked by a white cross. Extended emission to the northeast of the hot core is clearly visible in the lower $K$ transitions.}
 \label{Fig:6-5_intensities}
\end{figure*}

The observational data used in this paper were obtained as part of a 2-dimensional (2D) line survey of the Orion KL nebula, performed at the Institut de Radioastronomie Millim\'etrique (IRAM) 30m telescope in Pico Veleta, Spain. This mapping survey covers a $\sim$$2\arcmin\times2\arcmin$ region around Orion IRc2, with full spatial sampling and continuous frequency coverage across the 1~mm atmospheric window (from 200 to 282 GHz). A complete description of the 2D survey is presented in Marcelino et al. (\textit{in prep.}). We describe here only the details specific to the observations covering the methyl cyanide transitions analysed in this paper.

The observations presented here were obtained on 2008 February 16 ($J$=$12_K$--$11_K$), 2010 February 14 ($J$=$6_K$--$5_K$ and $13_K$--$12_K$), and 2012 January 22 ($14_K$--$13_K$). The EMIR single pixel heterodyne receivers were used for all observations except for those performed in 2008, for which the 9-pixel HERA receiver array was used. The $J$=$6_K$--$5_K$, $J$=$12_K$--$11_K$, $J$=$13_K$--$12_K$, and $J$=$14_K$--$13_K$ series of methyl cyanide $K$-ladder transitions were each observed using a single Local Oscillator setting, at frequencies of 109.983, 220.700, 239.100, and 258.000~GHz, respectively. The possible presence of emission arising from the image sideband was checked by using short wobbler-switching observations on the central position with a slightly different frequency for each setting. All maps were centred on the infrared continuum source IRc2 at $\alpha$=$5$h\,35m\,14.49s, $\delta$=$-5^\circ$\,22$\arcmin$\,29.3$\arcsec$ (J2000.0) and covered an area of 140$\times$140 arcsec$^2$ with map points separated by 4 arcsec. The observations were performed using the \textit{On-The-Fly} (OTF) mapping mode, scanning both in $\alpha$ and $\delta$, with position switching to an emission-free reference position at an offset $(-600\arcsec,0\arcsec)$ with respect to IRc2. The WILMA backend spectrometer was used for all observations, with a total bandwidth of 4 GHz (EMIR) and 1 GHz (HERA) and a spectral resolution of 2 MHz, corresponding to velocity resolutions of 5.4 to 2.3 km\,s$^{-1}$ at 110 and 258 GHz, respectively. Weather conditions were typically good winter conditions, with opacities $\sim$0.1--0.2 at 1 mm and 1--2 mm of precipitable water vapour (pwv), resulting in system temperatures of 200--300 K, except for the 2010 February period, when observations were performed with opacities of 0.3--0.4 and 5 mm of pwv. In this case, system temperatures of 150 K and 300--400 K were obtained at 110 and 239 GHz, respectively. Intensity calibration was performed using two absorbers at different temperatures and corrected for atmospheric attenuation using the Atmospheric Transmission Model \citep[ATM;][]{Cernicharo1985, Pardo2001}. The telescope half-power-beam-width (HPBW) was 22.3$\arcsec$ at 110~GHz, 11.1$\arcsec$ at 220~GHz, 10.3$\arcsec$ at 239~GHz, and 9.6$\arcsec$ at 258~GHz. The telescope pointing was checked every hour on strong nearby quasars and found to have errors of typically less than 3--4 arcsec. The flux accuracy was checked using repeated observations of the planet Mars and the estimated flux uncertainty is $\sim$25\%.

The data were processed using the IRAM GILDAS software package\footnote{See \texttt{http://www.iram.fr/IRAMFR/GILDAS} for more information about the GILDAS software package.}. Data reduction consisted of fitting and removing first order polynomial baselines, checking for image sideband contamination and emission from the reference position. The HERA data required further reduction analysis due to the different performance of each pixel in the array. Spectra from all pixels were averaged to obtain a uniform map spacing of 4$\arcsec$, taking into account the different flux calibration and internal pointing errors of each pixel (Marcelino et al.~\textit{in prep.}).


\section{Methyl cyanide emission maps}\label{Results:Emission-Maps}

\begin{figure*}
 \centering
 \includegraphics[width=17cm]{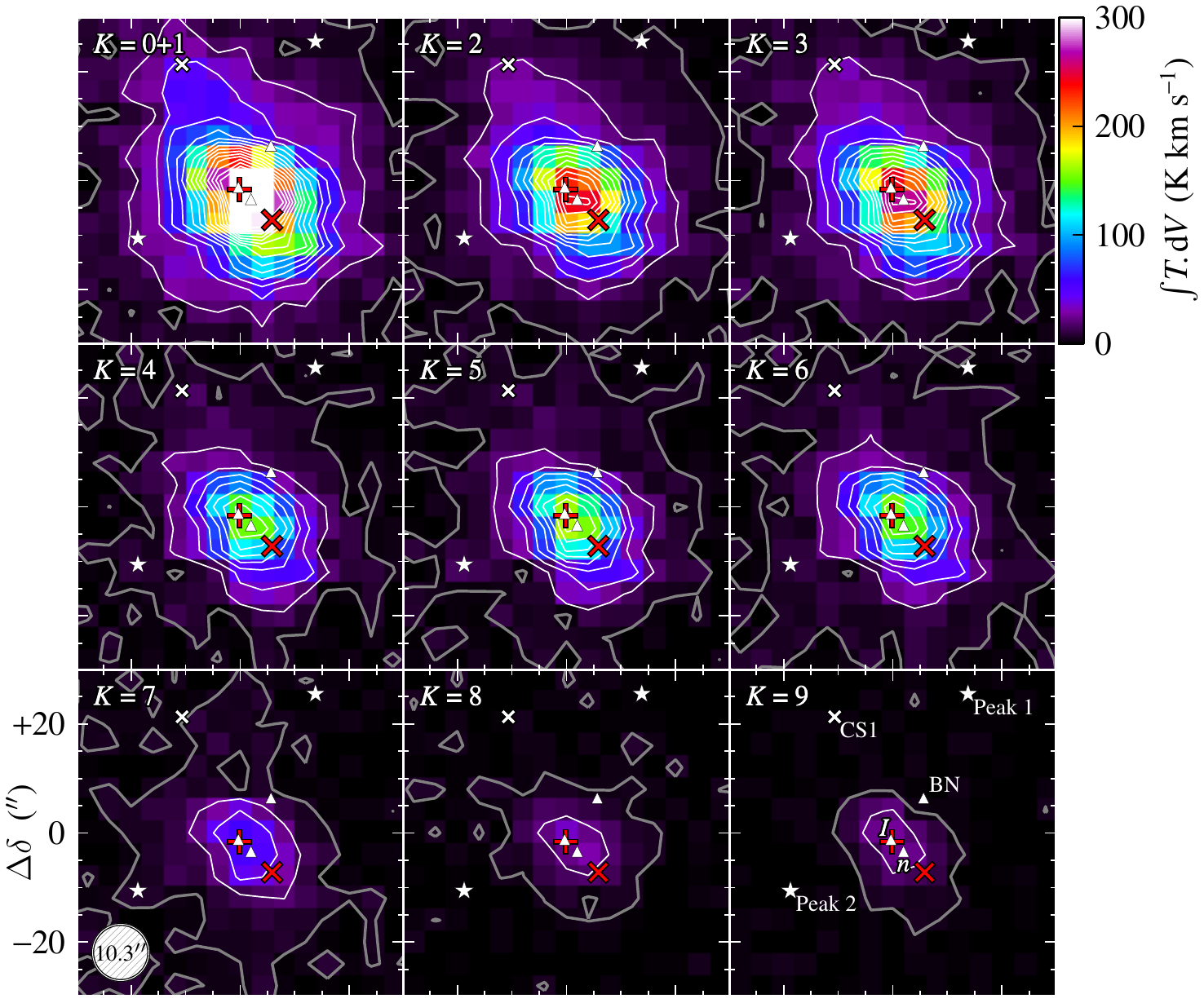}
 \caption{Integrated intensity maps of methyl cyanide $J$=$13_K$--$12_K$ emission across the Orion KL region. Contours are plotted at the 3-sigma level (4.8~K\,km\,s$^{-1}$, grey) and from 20 to 280 in 20~K\,km\,s$^{-1}$ intervals (white lines). The various sources indicated on the maps are as described in Fig.~\ref{Fig:6-5_intensities}.}
 \label{Fig:13-12_intensities}
\end{figure*}

Integrated intensity maps for the $J$=$6_K$--$5_K$ and $J$=$13_K$--$12_K$ sets of transitions are shown in Figs.~\ref{Fig:6-5_intensities} and \ref{Fig:13-12_intensities} (the $J$=$12_K$--$11_K$ and $14_K$--$13_K$ lines show similar distributions). The velocity ranges used to determine the integrated intensities were $[-12, +20]$ km\,s$^{-1}$ for the combined $K$=0$+$1 line emission, $[-2,+20]$ for the $K$=2 lines, to exclude emission from adjacent $K$-ladder lines, and $[-10,+25]$ for the higher $K$ lines, with velocity ranges adjusted slightly in some cases to avoid contamination from other species (see also Figs.~\ref{Fig:6-5_transition_labels} and \ref{Fig:12-11_13-12_14-13_transition_labels}). We note that the emission in the $J$=$12_K$--$11_K$ lines (not shown) has a noticeably smoother distribution than the $13_K$--$12_K$ and $14_K$--$13_K$ lines, despite the similar beam size. This is because the $12_K$--$11_K$ data were obtained with the HERA receiver array, whereas the other data were obtained with the EMIR single pixel receivers. Data obtained with the HERA array are more complicated to reduce due to the need to account for the different sensitivity and sideband rejection of each pixel (see details in the previous section) and small scale structure can be lost when averaging the spectra from the various receivers in the array, since minor pointing misalignments between pixels leads to ``blurring out'' of small features when they are averaged together. The $13_K$--$12_K$ observations were also obtained under poor weather conditions, leading to higher noise for these spectra.

The $K$-ladder transitions of methyl cyanide have upper levels spanning a wide range of energies, so the spatial extent of their emission show significant differences in morphology. In particular, the lower energy transitions typically show emission over a large expanse of the mapped area, whilst the high-$K$ lines arise in a compact region centred on the hot core, with some extension toward the southwest, suggesting an additional contribution from the compact ridge. Since the Einstein $A_{i\!j}$ coefficients do not drop dramatically with increasing $K$ in a given $K$-ladder (see Table~\ref{Table:Line-Properties}), the sudden decrease in size of the emitting region seen in the $K$=4 and higher lines compared to the lower energy transitions is due to a real change in the physical properties of the gas between the inner and outer parts of Orion KL. This implies that the central region around IRc2 must possess higher temperatures than the gas further out. Emission is detected in all transitions up to $K$=10, but becomes too weak to be detected in higher energy $K$-ladder lines beyond that. The elongated shape of the emission in the direction from northeast to southwest is consistent with that found in previous single-dish and interferometric observations of methyl cyanide \citep[e.g.,][]{Wilner1994, Wang2010}. However, we find the emission to be extended over a wider region -- more than 40 arcsec across -- compared to that found by interferometric observations, since such observations tend to filter out the extended emission on scales much larger than the synthesized beam size (\citealt{Wang2010}, for example, estimated that their interferometric observations missed up to 65\% of the flux in the more extended low-$K$ lines and filtered out structure on spatial scales greater than $\sim$14 arcsec). As such, the maps presented here provide an excellent dataset with which to study the emission properties of methyl cyanide over a much wider area surrounding Orion KL.

\begin{figure*}
 \centering
 \includegraphics[width=17cm]{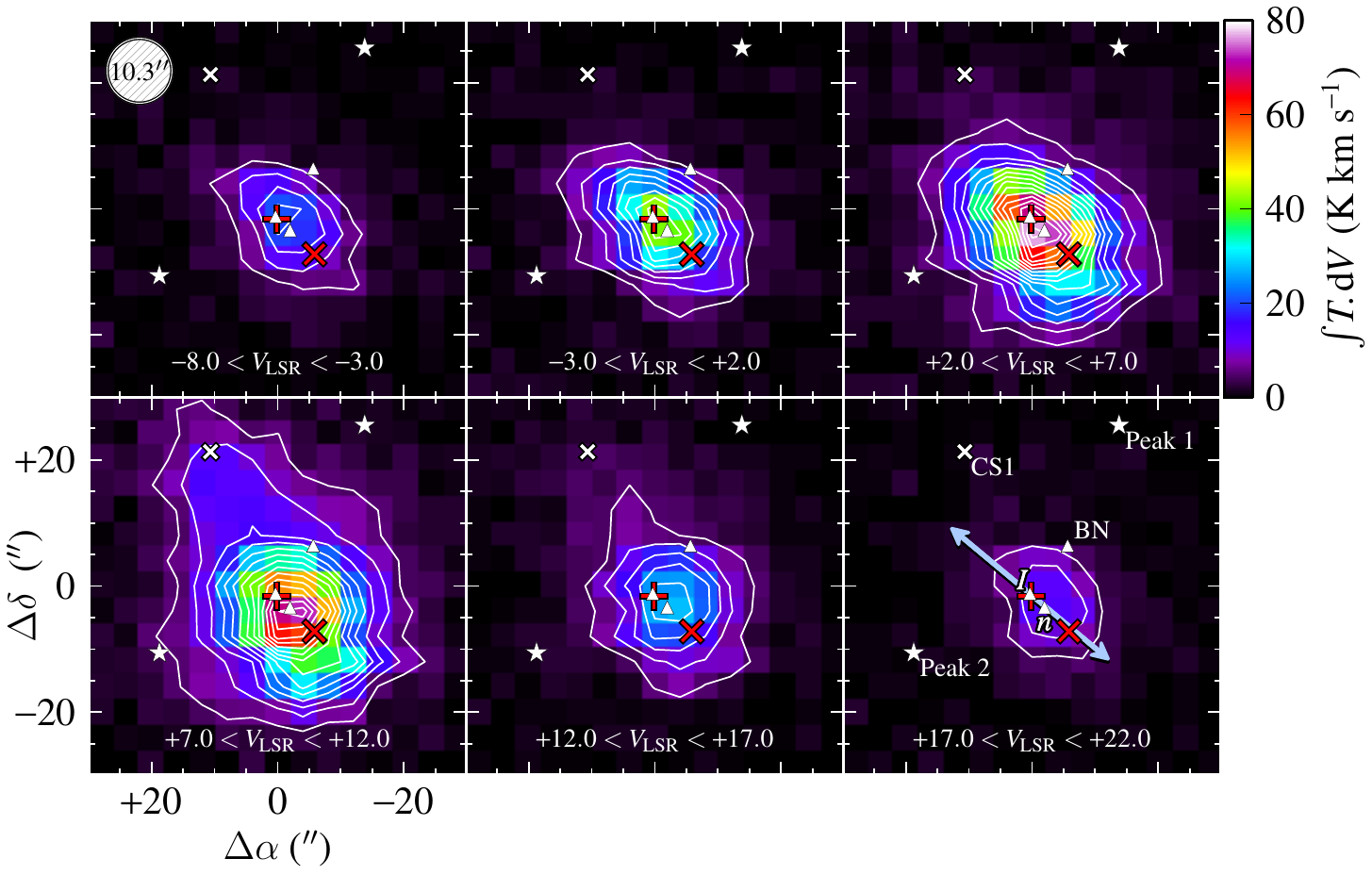}
 \caption{Channel maps of methyl cyanide $J$=$13_3$--$12_3$ emission integrated over the velocity ranges indicated on each panel. Contours are plotted from 5 to 70 in 5 K\,km\,s$^{-1}$ intervals. The various sources indicated on the maps are as described in Fig.~\ref{Fig:6-5_intensities}. The direction of the SiO outflow from source \textit{I} \citep[e.g.][]{Plambeck2009} is indicated by the blue arrow. The extended emission appears at velocities of 7--12 km\,s$^{-1}$, typical of the ridge.}
 \label{Fig:13-12_velocity_cuts}
\end{figure*}

\begin{figure}
 \resizebox{\hsize}{!}{\includegraphics{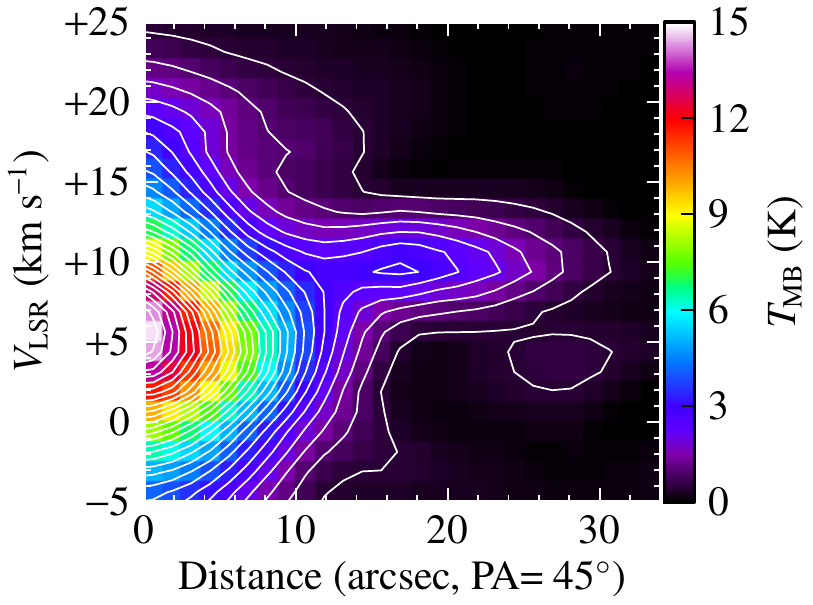}}
 \caption{Position-velocity diagram for the methyl cyanide $J$=$13_3$--$12_3$ emission along the northeast axis from IRc2. Distances are measured in the direction with position angle 45$^{\circ}$ from the map centre. Contours are plotted from 0.5 to 14 in 0.5 K intervals. A second emission peak is visible $\sim$17 arcsec to the northeast of IRc2, at the position $(+12\arcsec,+12\arcsec)$.}
 \label{Fig:13-12_pv_diagram}
\end{figure}

The kinematics of the CH$_{3}$CN emitting gas are revealed by channel maps of the emission integrated over adjacent velocity ranges, as shown in Fig.~\ref{Fig:13-12_velocity_cuts} for the $J$=$13_3$--$12_3$ transition. We choose to study the kinematics using this line because lower $K$-ladder transitions significantly overlap one another, whilst higher $K$-ladder transitions have upper state energies of $\sim$200 K or more and so do not emit in the cooler extended gas. Emission from this line is detected over LSR velocities from approximately \hbox{$-10$ to $+25$ km\,s$^{-1}$}. The broad velocity range of this emission suggests a contribution from gas in the plateau, a region associated with outflows generated by star formation within KL. The extremes of the blue- and red-shifted emission arise in a compact region roughly centred on the hot core and slightly elongated along the direction from northeast to southwest. This is consistent with observations of other molecules that trace the low-velocity outflow \citep{Plambeck1982, Genzel1989, Greenhill1998, Plambeck2009, Niederhofer2012}. The emission peaks at $\sim$6 km\,s$^{-1}$, consistent with the systemic velocity of the hot core, and is located approximately midway between the hot core and compact ridge, coinciding with the radio continuum source \textit{n} (indicated on Fig.~\ref{Fig:13-12_velocity_cuts}). Emission from the ridge components (both the compact and extended ridge) typically displays velocities in the range \hbox{$+8$ to $+12$ km\,s$^{-1}$}, and the methyl cyanide emission in this range (shown in the bottom-left panel of Fig.~\ref{Fig:13-12_velocity_cuts}) indeed peaks closer to the compact ridge position -- further evidence of emission from this component -- and extends over a much wider region around KL, therefore likely arising in the extended ridge.

\begin{figure}
 \resizebox{\hsize}{!}{\includegraphics{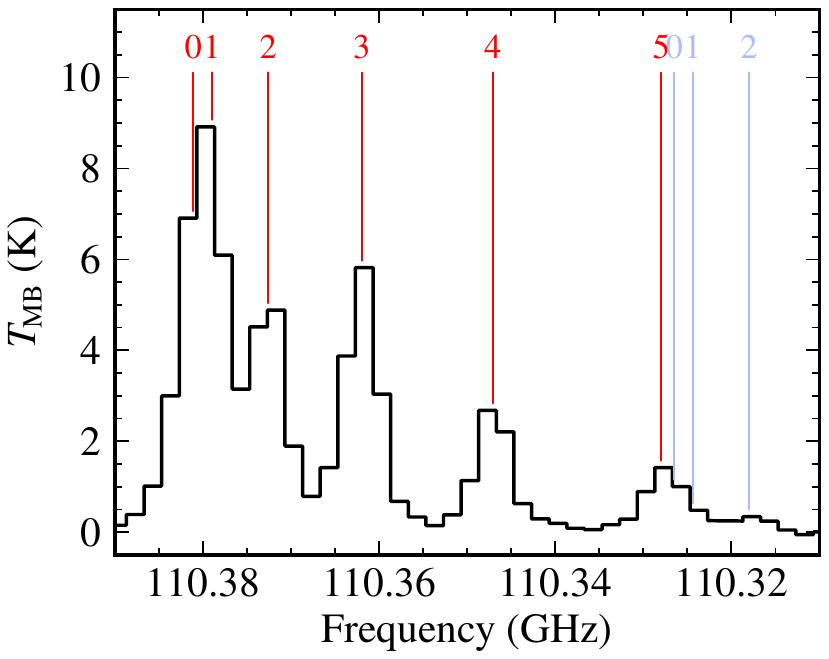}}
 \caption{The methyl cyanide $J$=$6_K$--$5_K$ spectrum observed toward IRc2 (the central position of the map). The $K$-ladder lines of the main isotopologue of methyl cyanide are labelled in red and those of CH$_3$$^{13}$CN are labelled in light blue. Note that the $K$=5 transition of methyl cyanide is contaminated by the CH$_3$$^{13}$CN $J$=$6_0$--$5_0$ and $J$=$6_1$--$5_1$ lines, as is the case for all $K$-ladder series.}
 \label{Fig:6-5_transition_labels}
\end{figure}

We also plot the position-velocity diagram for the same $J$=$13_3$--$12_3$ line along the northeast direction from IRc2 in Fig.~\ref{Fig:13-12_pv_diagram}. This direction follows the transition from the hot core into the quiescent ridge and the resulting diagram shows the shift from strong emission over a broad velocity range in the vicinity of IRc2 to weaker emission with narrow line width in the extended ridge. A second, minor emission peak occurs approximately 17 arcsec northeast of IRc2, corresponding to the position $(+12\arcsec,+12\arcsec)$. This peak appears at $\sim$10 km\,s$^{-1}$ and is not evident in the integrated intensity or channel maps, since the line is narrow and its integrated intensity therefore does not rise significantly above that of its surroundings. There is also evidence for weak emission displaying broad line wings in the region just before this emission peak. Taken together, we argue that these features show an abrupt change in the gas kinematics, where the low-velocity outflow meets the quiescent gas in the extended ridge. As we discuss later, this region also has physical properties that suggest it to be a distinct feature with respect to the surrounding gas.

The rms noise sensitivity achieved in these maps is typically $T_\mathrm{rms} \! \approx \! 50$--75~mK for the $J$=$6_K$--$5_K$ spectra and $\approx$100--170~mK for the higher frequency transitions. Given the rich line density produced by Orion KL, this high sensitivity means that many emission lines from other species are detected in the observed spectra, in addition to those of CH$_3$CN. To illustrate this, Figs.~\ref{Fig:6-5_transition_labels} and \ref{Fig:12-11_13-12_14-13_transition_labels} show the observed spectra for the $J$=$6_K$--$5_K$ up to $J$=$14_K$--$13_K$ transitions obtained towards IRc2, the central position of the maps. The observed spectra at all positions have frequency ranges much wider than those shown here and line-free regions of the spectra on both sides of the methyl cyanide $K$-ladder series were used to determine appropriate baselines. The $K$-ladder lines of methyl cyanide, its isotopologue CH$_3$$^{13}$CN, and detectable lines from a variety of other species are indicated on the spectra. In all four cases, the $K$=0 and 1 lines significantly overlap one another, leading to difficulties in fitting their line shapes, as discussed in the next section. In addition, the $K$=5 transition of each series of CH$_3$CN $K$-ladder lines is contaminated by emission from the $K$=0 and 1 transitions of CH$_3$$^{13}$CN. Though weak, these contaminating lines nevertheless contribute to the total flux, making the determination of the $K$=5 line intensity somewhat uncertain (also discussed in the next section). Apart from the contamination from its isotopologue, the observed lines of methyl cyanide are mostly unblended, with the exception of the $J$=$12_9$--$11_9$ line, which is completely covered by the $^{13}$CO 2--1 line (see Fig.~\ref{Fig:12-11_13-12_14-13_transition_labels}, top panel) and is therefore excluded from the analysis in this paper. Fitting the methyl cyanide lines with Gaussian profiles therefore provides reasonably accurate integrated intensities, as described in the following section, and allows the kinematics of the emitting gas to be studied in more detail.

\begin{figure*}
 \centering
 \includegraphics[width=17cm]{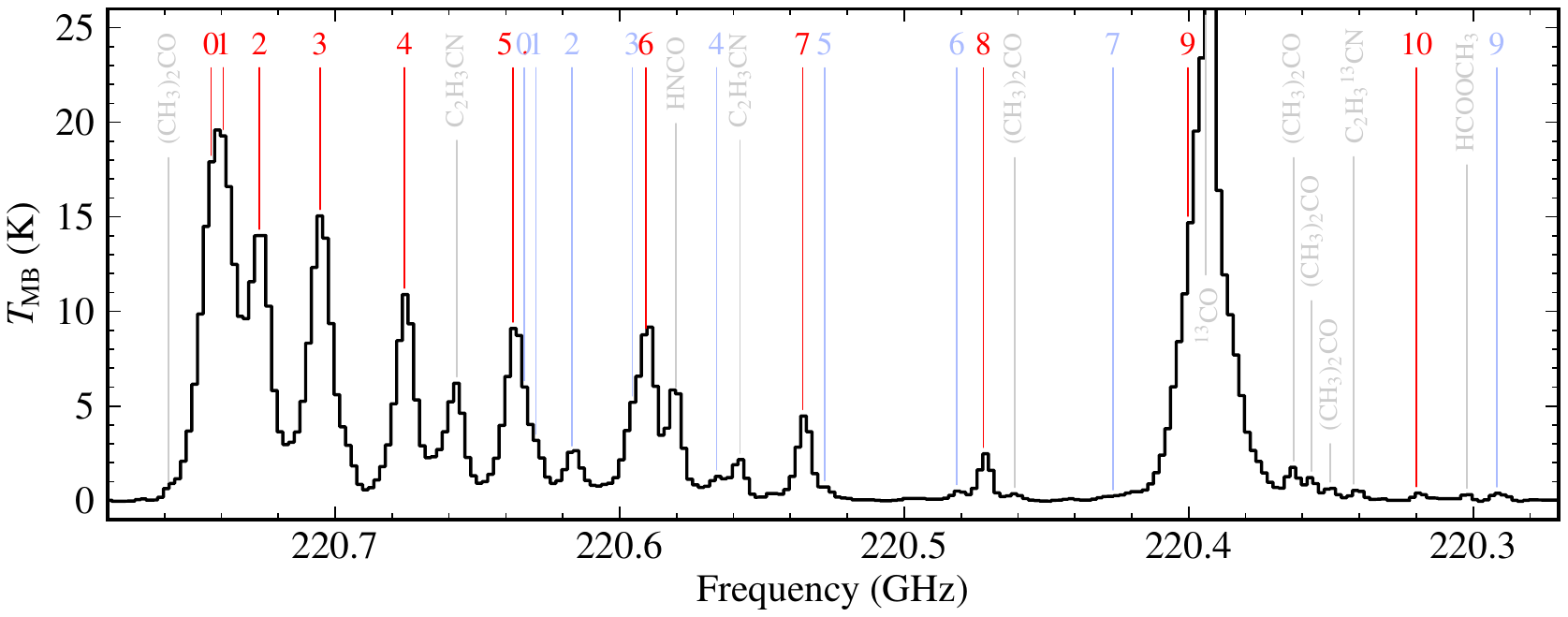}
 \includegraphics[width=17cm]{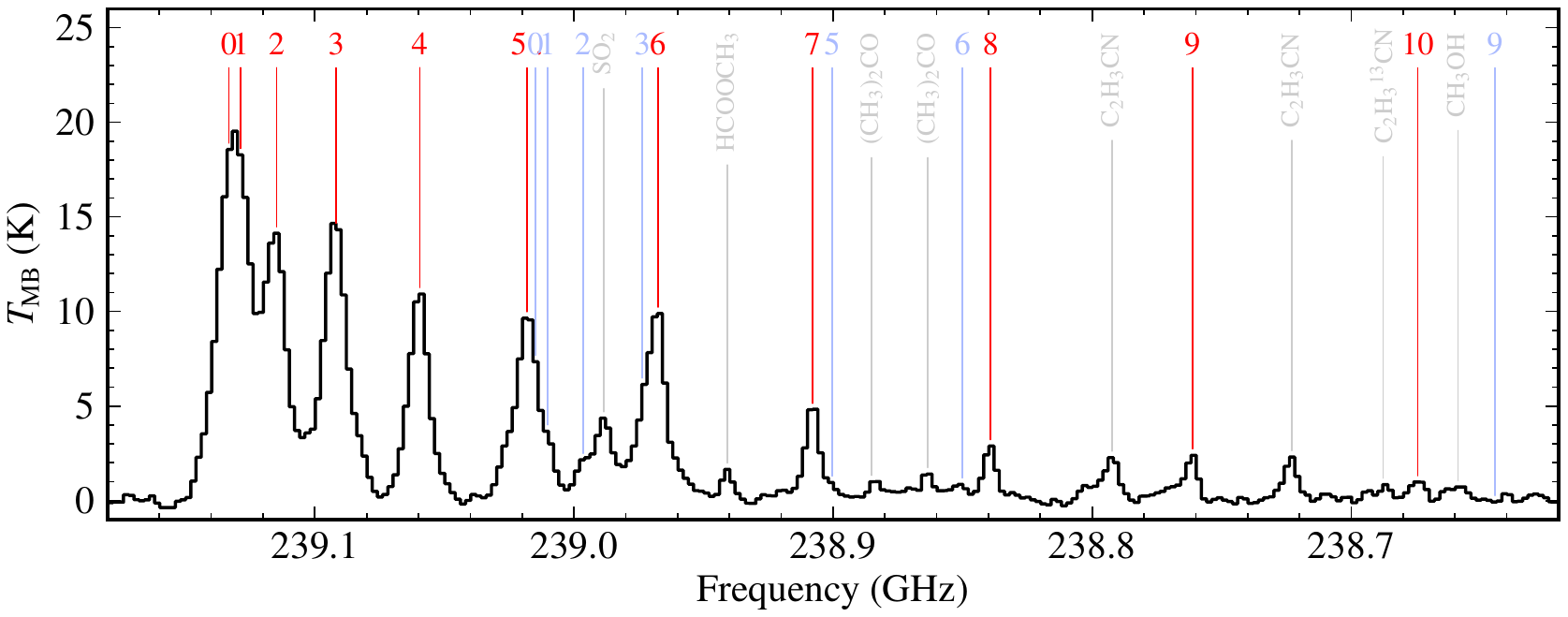}
 \includegraphics[width=17cm]{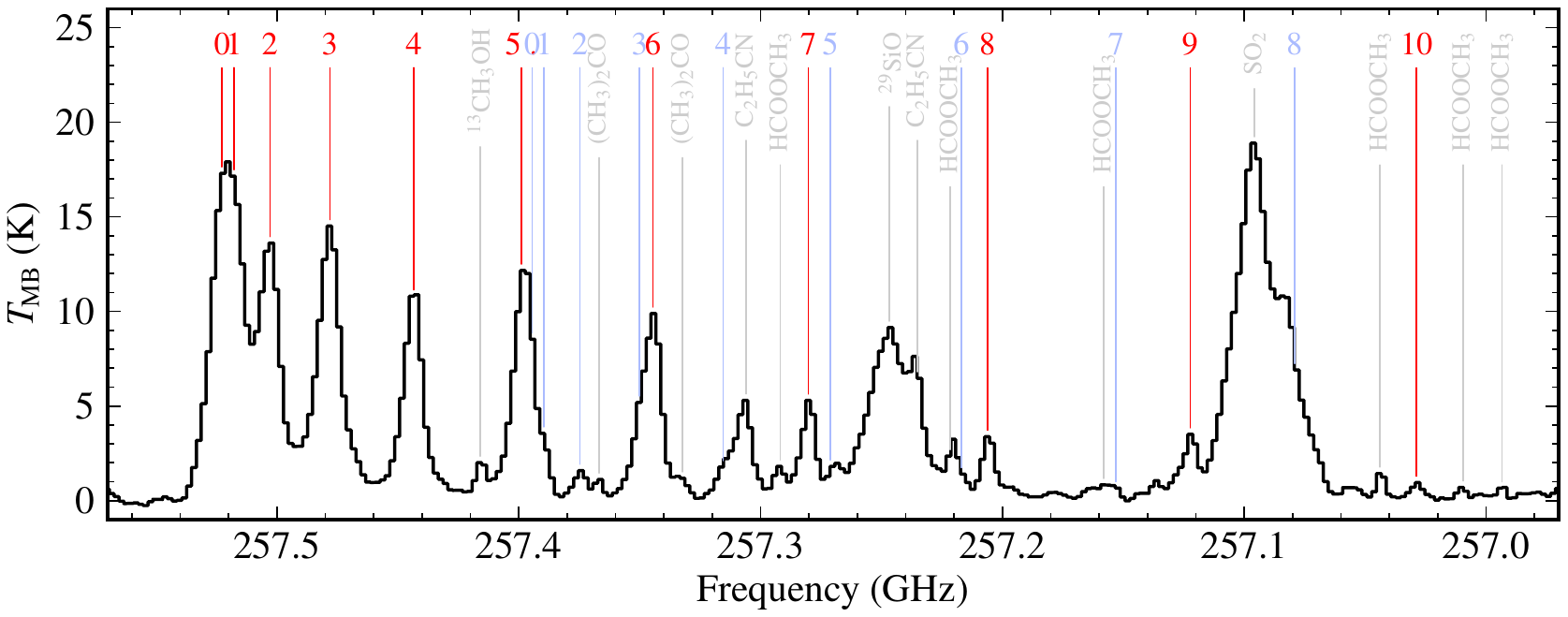}
 \caption{The methyl cyanide $J$=$12_K$--$11_K$ (\textit{top}), $13_K$--$12_K$ (\textit{middle}), and $14_K$--$13_K$ (\textit{bottom}) spectra observed toward IRc2, identifying the various lines detected in these frequency ranges. The $K$-ladder lines of the main isotopologue of methyl cyanide are labelled in red, those of the CH$_3$$^{13}$CN isotopologue are labelled in light blue, and emission lines from other species are indicated in grey. Note that the $J$=$12_9$--$11_9$ transition suffers strong contamination from the $^{13}$CO $J$=2--1 line.}
 \label{Fig:12-11_13-12_14-13_transition_labels}
\end{figure*}


\section{Determining temperatures and column densities}\label{Analysis}

Given its regular series of $K$-ladder transitions that are closely grouped in frequency and span a wide range of upper level energies, methyl cyanide is an excellent molecule with which to determine rotational temperatures by use of rotation or population diagram analysis \citep{Turner1991, Goldsmith1999}. Several assumptions are made when using these techniques, foremost of which is the assumption that the level populations are thermalised and so are governed by collisional excitation (as opposed to radiative excitation). In this section we describe the methods used to fit Gaussian profiles to the methyl cyanide lines at each position in the maps and the subsequent rotation and population diagram analyses based on the fitted line intensities. A more advanced analysis made by fitting LVG models to the observed line intensities is also described.


\subsection{Gaussian fits to the emission lines}\label{Analysis:Gaussian-Fits}

Integrated intensities for the methyl cyanide emission lines were determined at each map position by simultaneously fitting Gaussian profiles to the $K$=0--10 lines of each observed set of $K$-ladder transitions (with the exception of the $J$=6--5 transitions, for which all six $K$=0--5 lines were fitted). Emission from the $K$=0 and $K$=1 transitions frequently overlaps, particularly for the broader line widths found in the Orion hot core. Disentangling the integrated intensities produced by these two transitions is therefore difficult, especially since these lines can often exhibit strong opacities. Assuming that all lines arise from the same gas within a given map position, we first constrained the fits by requiring that all transitions within a given $K$-ladder have identical line widths and central LSR velocities. The Gaussian profile fits for the $K$=0 and $K$=1 lines were also constrained by requiring their fluxes to be equal, a reasonable approximation since their upper state energies and line strengths are comparable. In our fitting routine we allowed for two kinematic components -- one narrow, with allowed line widths of 3--10 km\,s$^{-1}$ (FWHM), and one broad, with line widths of 10--25 km\,s$^{-1}$ -- in order to adequately fit the line profiles at some map positions. The decision to include one or two kinematic components at each position was made based on their relative goodness of fit using the Bayesian information criterion \citep[BIC;][]{Schwarz1978}. This criterion introduces a penalty term for the number of model parameters in the goodness of fit to prevent overfitting. If the BIC value was lower for a fit with a single kinematic component at a given position, this fit was adopted instead of the two-component fit. Generally, we found that two kinematic components were needed to fit the line emission in the central region surrounding the hot core and compact ridge, but that one was sufficient for the majority of positions covering the more quiescent gas. The unconstrained parameters were allowed to vary freely from position to position within the maps, however some degree of consistency was imposed between adjacent map positions by using the best-fit parameters derived at one position as the initial guess values for fitting the lines at the next adjacent position. The map spectra were fitted in sequence, starting at the central position and moving to each neighbouring position following an outward spiral pattern.

\begin{figure}
 \resizebox{\hsize}{!}{\includegraphics{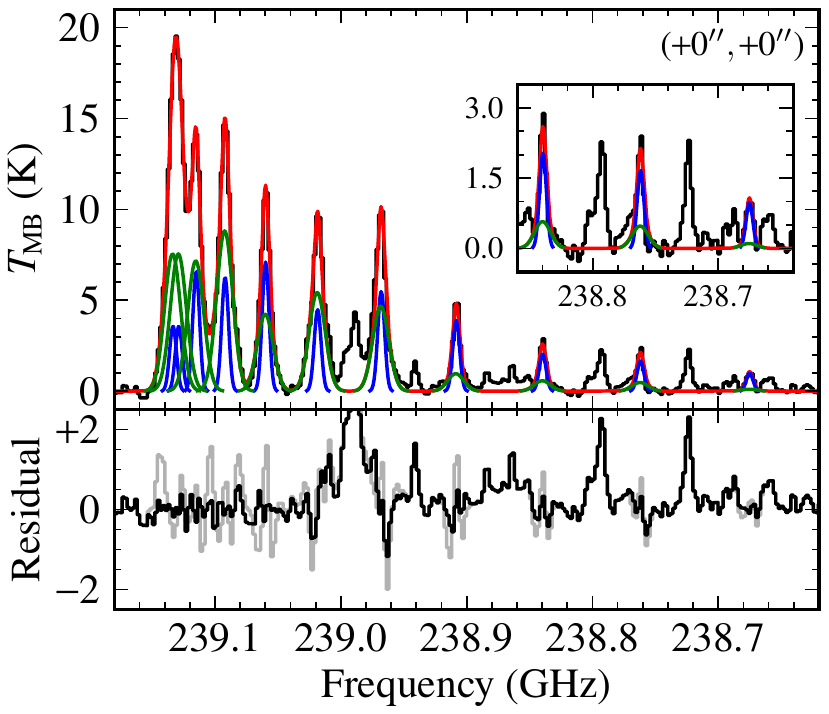}}
 \caption{Best-fit Gaussian profiles for the $J$=13--12, $K$=0--10 methyl cyanide lines at the central position in the Orion KL map. Identical line widths and central velocities are assumed for all $K$-ladder transitions within a given kinematic component (see Sect.~\ref{Analysis:Gaussian-Fits} for details). Two kinematic components (one narrow, shown in blue, and one broad, shown in green) have been fitted while allowing their respective line widths and central velocities to vary. The combined fit from all Gaussian profiles is shown in red. The inset panel shows a zoom of the region around the $K$=8--10 lines. The residual for the two-component fit is shown in black in the bottom panel; the residual for the best one-component fit is shown in grey for comparison. The residual for the fit using one kinematic component is noticeably worse than that with two.}
 \label{Fig:13-12_gaussian-fits-centre}
\end{figure}
\begin{figure}
 \resizebox{\hsize}{!}{\includegraphics{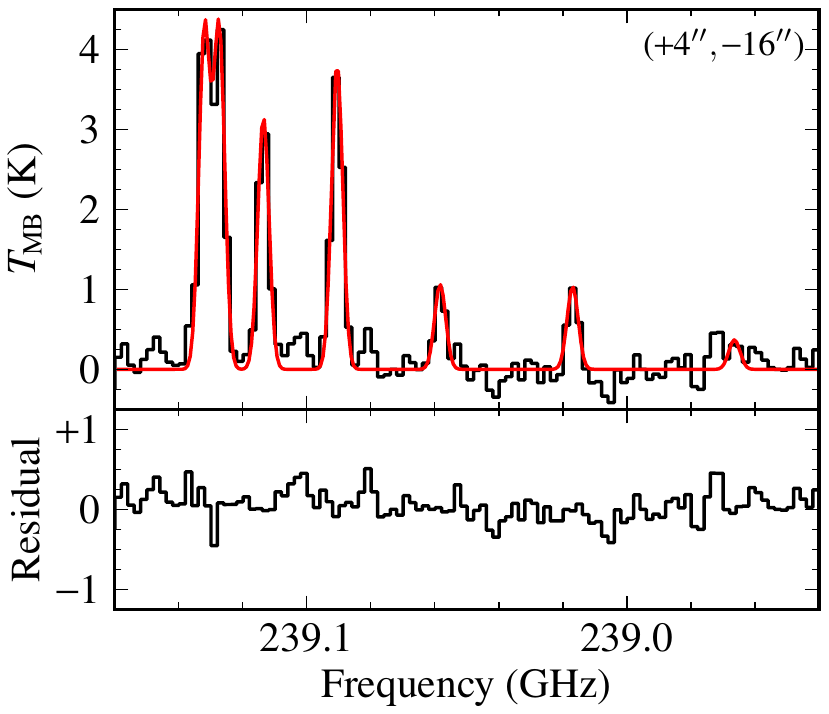}}
 \caption{Best-fit Gaussian profiles for the $J$=13--12, $K$=0--6 methyl cyanide lines at the offset position $(+4\arcsec,-16\arcsec)$ in the Orion KL map. Identical line widths and central velocities were assumed for all $K$-ladder transitions within a given kinematic component (see Sect.~\ref{Analysis:Gaussian-Fits}). The $K$-ladder lines above $K$=6 were not detected at this position, so are excluded from the figure. Only one kinematic component (shown in red) is necessary to obtain a good fit in this case.}
 \label{Fig:13-12_gaussian-fits-offset}
\end{figure}

The fit to the $J$=$13_K$--$12_K$ spectrum at the central position of IRc2 is shown in Fig.~\ref{Fig:13-12_gaussian-fits-centre}, illustrating that two kinematic components, one narrow ($\sim$8 km\,s$^{-1}$) and one broad ($\sim$19 km\,s$^{-1}$), both with comparable central velocities (5.7 versus 5.2 km\,s$^{-1}$), fit remarkably well the observed spectrum. Note that a fit with only one component yields residuals twice as large. These kinematic signatures can be assigned to emission arising in the hot core and plateau components, respectively. The narrower hot core component is seen to dominate in the high energy $K$-ladder lines (to the right of the spectrum), consistent with the presence of higher temperatures in this source. In contrast, more quiescent regions away from the centre require only one kinematic component to fit the line profiles, as shown in Fig.~\ref{Fig:13-12_gaussian-fits-offset}, which have kinematic properties similar to those of the extended ridge (narrow lines with FWHM of $\sim$5 km\,s$^{-1}$ at central velocities of $\sim$9 km\,s$^{-1}$).

\begin{figure*}
 \centering
 \includegraphics[width=17cm]{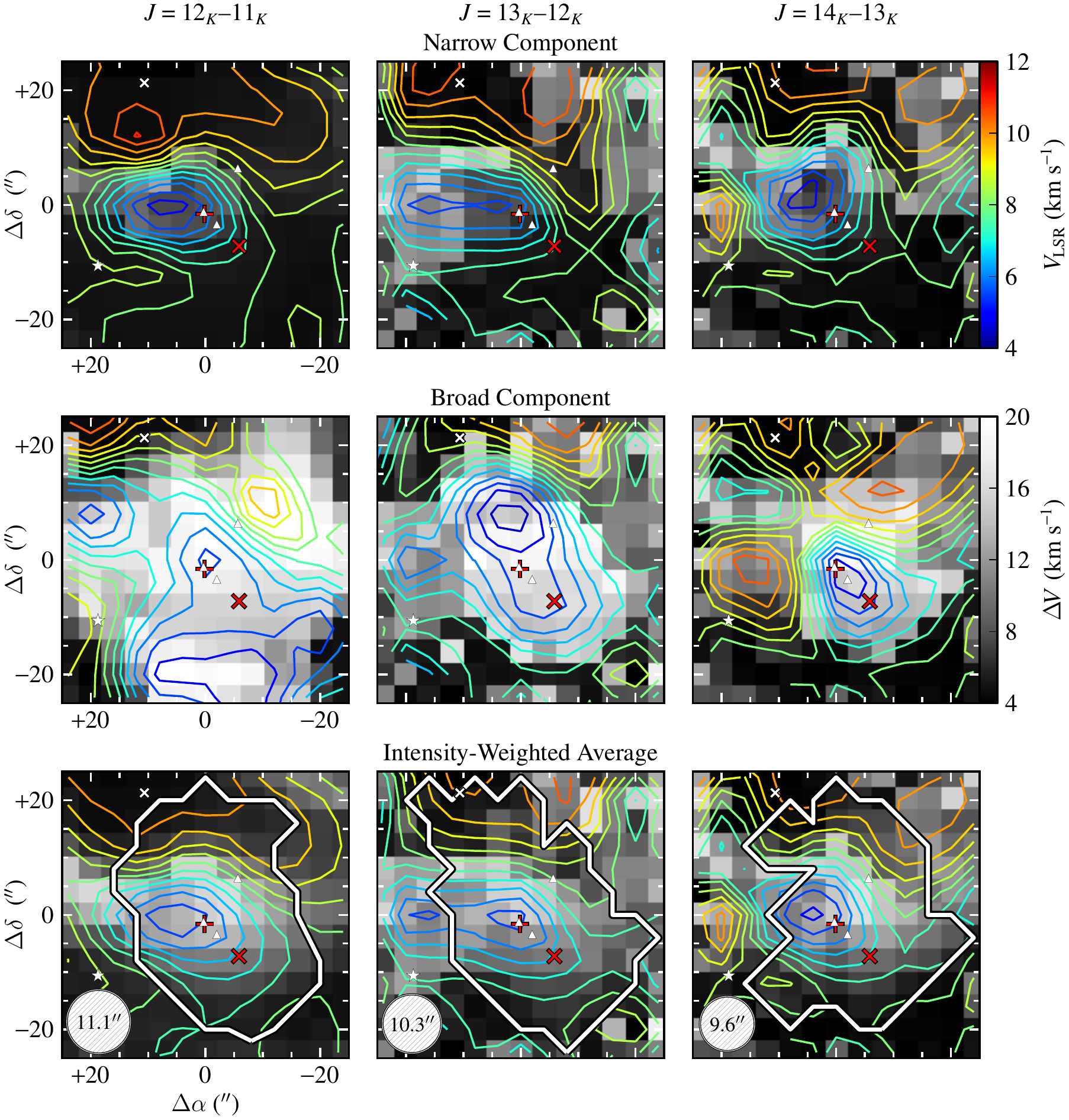}
 \caption{Distribution of the LSR velocity ($V_\mathrm{LSR}$, coloured contours) and line width ($\Delta V$, greyscale maps) of methyl cyanide $J$=$12_K$--$11_K$ (\textit{left}), $J$=$13_K$--$12_K$ (\textit{centre}), and $J$=$14_K$--$13_K$ (\textit{right}) emission across the Orion KL region, as determined from simultaneous Gaussian fits to the $K$=\,0\,--\,10 lines (see Sect.~\ref{Analysis:Gaussian-Fits} for details). The three rows show the kinematic properties from the two-component fits for the narrow component (\textit{top}), the broad component (\textit{middle}), and the intensity-weighted average of the two (\textit{bottom}). Thick white contours on the bottom panels indicate the regions where two kinematic components were required to obtain a good fit to the line profiles. The various sources indicated are as described in Fig.~\ref{Fig:6-5_intensities}.}
 \label{Fig:Kinematics}
\end{figure*}

Figure~\ref{Fig:Kinematics} shows the kinematic properties (LSR velocity $V_\mathrm{LSR}$ and FWHM line width $\Delta V$) of the best-fit Gaussian profiles for the $J$=$12_K$--$11_K$, $J$=$13_K$--$12_K$, and $J$=$14_K$--$13_K$ sets of $K$-ladder transitions across the Orion KL region. Separate panels show the properties of the narrow component, broad component and the intensity-weighted average properties from these fits. The extended CH$_3$CN emission shows a sharp north-south velocity gradient around IRc2, in addition to the more gradual velocity gradient over larger scales previously observed in its $J$=$5_K$--$4_K$ line emission by \citet{Wilner1994} and in other molecules that trace the extended ridge \citep[e.g.,][]{Vogel1984, Womack1991, Ungerechts1997}. Since the most blue-shifted emission is extending to the northeast of the hot core, much like the blue-shifted SiO emission studied by \citet{Plambeck2009}, and exhibits broader line widths, this emission is most likely associated with the low-velocity outflow from the KL region.

There is an additional emission feature displaying broad line widths ($\gtrsim$16 km\,s$^{-1}$) that extends toward the eastern edge of the maps. The emission in this outer region is generally quite weak, so the derived kinematic properties are less reliable and may be the result of fitting baseline ripples; however, the fact that we consistently see the same region of broad line width emission in all sets of $K$-ladder transitions suggests that it could be a real feature. If so, this region may be associated with the high-velocity outflow, as traced by the bright H$_2$ emission stretching northwest and east of Orion KL \citep{Beckwith1978} and the molecular ``fingers'' \citep{Taylor1984, Allen1993} that may have originated from a prior explosive event, as proposed by \citet{Bally2005}, \citet{Zapata2009, Zapata2011}, \citet{Niederhofer2012}, and \citet{Nissen2012}.

In the following analysis, at positions where two Gaussian components were required to fit the line profiles, we use the intensity-weighted average line width and the total integrated intensity from both components. This is clearly a simplifying assumption, and in the central region of the maps, where various distinct sources contribute to the emission within the beam, deriving physical properties from a single-component analysis can only yield average conditions for the region, masking the more complex underlying structure. However, the focus of this paper is on the conditions within the extended gas, outside of the hot core and away from IRc2. In the regions further from the centre, a single kinematic component typically dominates the line intensity, with little need for a second component to fit the line profile. Due to the limited spectral resolution of the data (2--5 km\,s$^{-1}$) and the weaker emission from the more extended gas, we have found that treating the two components separately can lead to inconsistent and sometimes unphysical results. We therefore consider the intensity-weighted average line width of the fitted profiles (which is generally dominated by only one of the components), to be the more reliable estimate in the weaker extended emission.

The crowded region encompassing the hot core and compact ridge deserves a more detailed analysis, which is beyond the scope of this paper. Instead, such an analysis has been carried out as part of a separate, dedicated study of the central position that includes all methyl cyanide lines detected in both IRAM and \textit{Herschel}-HIFI spectral line surveys of the source \citep{Crockett2014}. This study treats the distinct kinematic components separately and is therefore better able to discuss the detailed properties of the central part of Orion KL.

We assume that the total uncertainty on the integrated line intensity is given by the sum of the uncertainty from the Gaussian fit and the estimated flux uncertainty in the observation, added in quadrature. An additional 25\% uncertainty is assumed for the line intensity of each $K$=5 transition to account for contamination from the overlapping CH$_3$$^{13}$CN lines (see Sect.~\ref{Results:Emission-Maps}). This represents an upper limit to the additional emission contributed by these lines, since they are expected to be weak.


\subsection{Constructing rotation diagrams}\label{Analysis:Rotation-Diagrams}

Estimates for the excitation temperature and column density of a molecular species can be determined via rotation diagram analysis \citep[e.g.,][]{Turner1991}, by assuming that its level populations are in local thermodynamic equilibrium (LTE), that the emission lines are optically thin, and that the emission arises in a source of uniform density and temperature, with no appreciable background radiation. The details of this analysis technique are described in Appendix~\ref{Appendix:Rotation-Diagrams}.

We have performed this analysis and constructed rotation diagrams for the methyl cyanide emission lines at all positions in our maps, thereby obtaining the distribution of rotational temperatures and total column densities across the observed region. The relevant spectroscopic quantities for our observed transitions of methyl cyanide have been taken from the Cologne Database for Molecular Spectroscopy \citep[CDMS;][]{Muller2005} and are listed in Table~\ref{Table:Line-Properties}. As previously discussed, at positions where two kinematic components were necessary to reproduce the line profiles, the total integrated intensity from both components is used in the rotation diagram analysis. Since the majority of the extended emission displays line profiles that are clearly dominated by one kinematic component, it is generally the case that this approach yields the properties for the dominant component. However, for the central region where multiple kinematic components contribute significantly to the total emission, this approach yields the average properties at these positions. This analysis has been performed separately for each set of $K$-ladder transitions. The inferred column densities are beam-averaged and do not account for the possible dilution of emission from unresolved clumps. Therefore, the column densities obtained with this method represent lower limits. For each set of $K$-ladder transitions, all the lines are observed simultaneously at one frequency setting and hence with the same beam size, so beam dilution does not affect the relative line strengths. The results from this analysis are presented and discussed in detail in Sect.~\ref{Results:Rotation-Diagrams}.


\subsection{Accounting for line opacity and beam dilution}\label{Analysis:Opacity-Dilution}

The observed line intensities are further reduced by line opacity when the column of emitting material is large, and beam dilution when the emitting region is smaller than the solid angle subtended by the telescope beam. In the case of methyl cyanide, the issue of opacity can be especially important for the low-$K$ transitions. Both line opacity and beam dilution are accounted for by the population diagram technique (as distinct from rotation diagrams), described in detail in Appendix~\ref{Appendix:Population-Diagrams}, which we use to further constrain the properties of the emitting gas at each position in the maps. In our analysis, we consider a region of parameter space covering rotational temperatures in the range 10\,--\,800~K, total column densities $N_\mathrm{tot}$\,=\,$10^{14}$\,--\,$10^{18}$~cm$^{-2}$, and beam dilution factors $f_\mathrm{beam}$\,=\,$10^{-2}$\,--\,$10^{0}$. The best-fit parameters obtained from this analysis are presented and discussed in Sect.~\ref{Results:Population-Diagrams}.


\subsection{LVG model fits}\label{Analysis:LVG-Model-Fits}

Population diagrams offer a powerful tool for determining rotational temperatures and analysing the excitation of a molecule, however, the relation to kinetic temperature is always dependent upon the assumption of local thermodynamic equilibrium. While this may be a reasonable assumption for the densest parts of Orion KL, the more tenuous gas in the extended cloud is likely to be below the critical densities needed to excite the transitions of methyl cyanide studied here ($n_\mathrm{crit} \!\sim\! 10^7$ cm$^{-3}$ at 100 K for the $J$=$12_K$--$11_K$ and $13_K$--$12_K$ transitions), leading to subthermally excited states. Furthermore, the opacities of the low-$K$ transitions of CH$_3$CN are known to be high, leading to a flattening of the slope in the population diagram and resulting in rotational excitation temperatures and total column densities that are over- and under-predicted, respectively.

We therefore also consider more sophisticated models of the emission; specifically, radiative transfer models that use the local velocity gradient (LVG) approximation to simplify the line transfer problem. By fitting LVG model predictions to the observed line intensities, it is possible to derive the kinetic temperature $T_\mathrm{kin}$, volume number density of molecular hydrogen $n(\mathrm{H_2})$, total column density of methyl cyanide $N_\mathrm{tot}$, and beam filling factor of the emitting region $f_\mathrm{beam}$. In order to achieve this for each position in our maps, we have run a grid of LVG models to predict line intensities for the observed methyl cyanide transitions over a region of parameter space that spans $T_\mathrm{kin}$=\,50\,--\,500 K, $n(\mathrm{H_2})$\,=\,$10^{4}$\,--\,$10^{8}$ cm$^{-3}$, $N_\mathrm{tot}$\,=\,$10^{14}$\,--\,$10^{18}$ cm$^{-2}$, and $f_\mathrm{beam}$\,=\,$10^{-2}$\,--\,$10^{0}$. We use the Madex LVG code \citep{Cernicharo2012} to perform these calculations, adopting the collisional excitation rates determined by \citet{Green1986} and extrapolated to higher temperatures.

Best-fit models for each map position were then determined by performing a $\chi^{2}$ minimisation over the entire grid of models, comparing the predicted integrated line intensities to those obtained by the Gaussian fits to the observed spectra (summing the intensities of both fitted components at positions where two components were used; see Sect.~\ref{Analysis:Gaussian-Fits} for details). As with the population diagram analysis, we determine the reduced chi-squared for a given set of observed and model line intensities using equation~\ref{Equation:Chi-Squared} (see Appendix~\ref{Appendix:Population-Diagrams} for details). The best-fit parameters are presented and discussed in Sect.~\ref{Results:LVG-Models} and \ref{Results:Combined-Models}.


\subsection{Limitations of the analysis methods}\label{Analysis:Limitations}

The three techniques discussed above to derive physical properties from the observed methyl cyanide lines are all subject to certain limitations and make assumptions about the emitting gas. The rotation diagram method yields rotational temperatures, not kinetic temperatures, with the two being equal only when the gas is sufficiently dense that collisions dominate its excitation. In addition, the gas is assumed to be optically thin and that the source fills the telescope beam, thus the derived column densities are beam-averaged values. The central region of Orion KL is sufficiently dense that methyl cyanide is likely to be thermalised, so that the derived rotational temperatures are reasonably close to the kinetic temperatures. However, the expected column densities are high enough in this region that line opacities become non-negligible, more so in the low-lying $K$-ladder lines than in the higher energy transitions. This reduces the observed intensities of these low-$K$ lines, leading to lower values being derived for their upper state column densities and a ``flattening'' of the slope in the resulting rotation diagram, which yields higher rotational temperatures.

\begin{figure*}
 \centering
 \resizebox{\hsize}{!}{\includegraphics{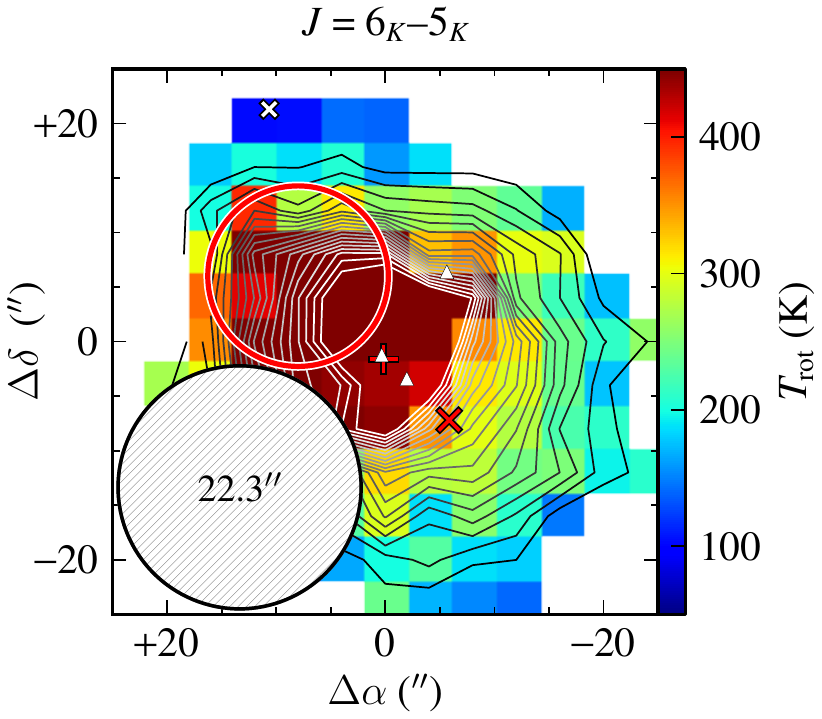}\hspace{5mm}
 \includegraphics{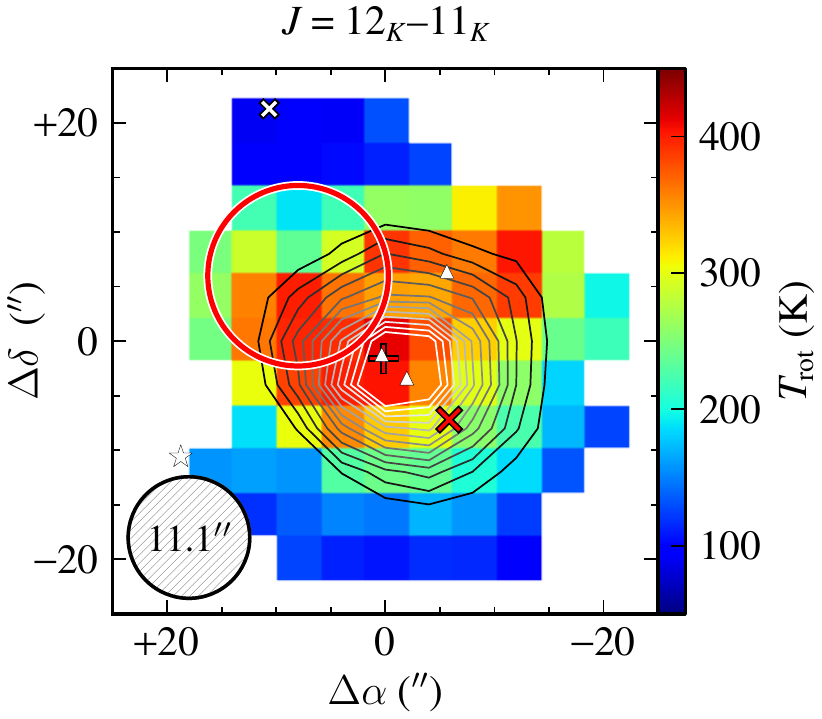}}\\
 \vspace{5mm}
 \resizebox{\hsize}{!}{\includegraphics{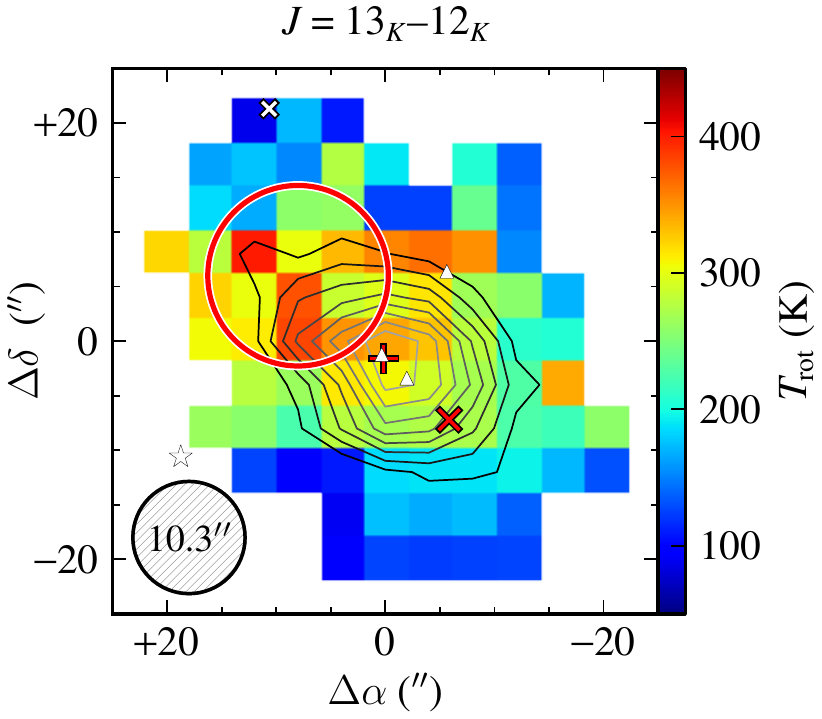}\hspace{5mm}
 \includegraphics{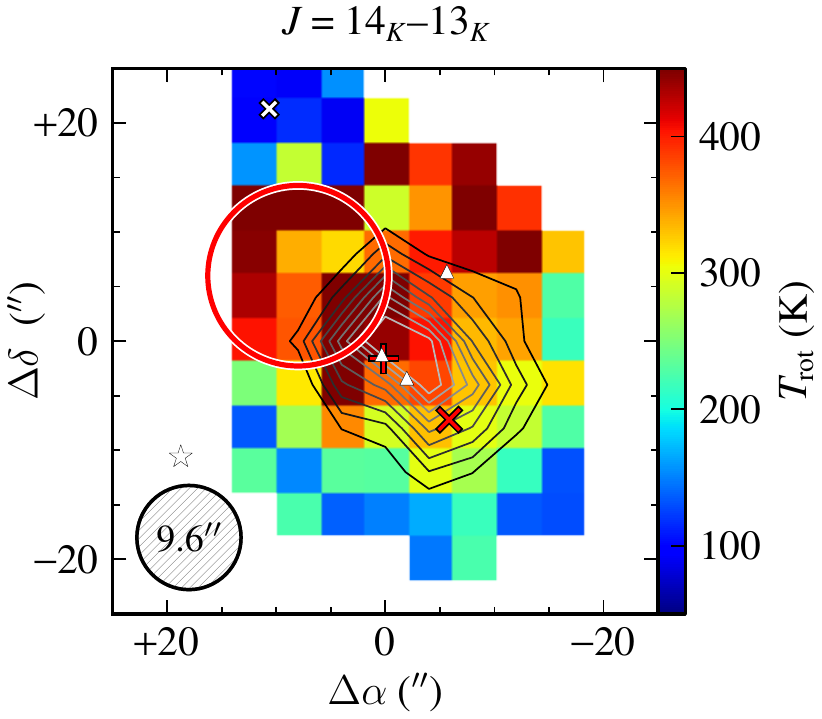}}
 \caption{Results of the rotation diagram analysis assuming optically thin emission. Maps of methyl cyanide rotational temperature ($T_\mathrm{rot}$, colour scale) and total beam-averaged column density ($N_\mathrm{tot}$, overlaid contours) across the Orion KL region. Both quantities are determined from rotation diagram analyses of the $J$=$6_K$--$5_K$, $J$=$12_K$--$11_K$, $J$=$13_K$--$12_K$, and $J$=$14_K$--$13_K$ line emission at each map position, neglecting the effects of line opacity and beam dilution (see Sect.~\ref{Analysis:Rotation-Diagrams} for details). Column density contours are plotted at 0.5$\times$10$^{15}$ cm$^{-2}$ intervals, starting at 1.0$\times$10$^{15}$ cm$^{-2}$ (black lines). The various sources indicated are as described in Fig.~\ref{Fig:6-5_intensities}, with the approximate location and extent of the hot zone seen to the northeast of IRc2 also marked by a red circle on each panel (defined in Sect.~\ref{Results:Temperature-Maps}).}
 \label{Fig:RT_Trot_Ntot_maps}
\end{figure*}

Population diagram analysis including the effects of line opacity and beam dilution can account for optically thick emission and unresolved source sizes. However, the assumption of LTE still remains, and this method cannot provide an indication of the volume density of the emitting gas. There is also significant degeneracy between the total column density and the source dilution factor, since one can increase to compensate the other. The change in line opacity that comes with increasing column density can help to break this degeneracy somewhat, but not completely. The slope of the population diagram is also linked to the line opacity, introducing some degeneracy in the temperature-column density plane of parameter space (since the opacity is governed by both), though less so than the coupling between total column and dilution factor.

Finally, LVG model fits to the observed line intensities can provide additional constraints on the volume density of the gas, though in the case of Orion KL, where densities are high enough to thermalise the population levels, these models may only provide lower limits for the density. LVG models also suffer the same degeneracies as population diagram techniques, though by combining data from various sets of $K$-ladder transitions observed with different beam sizes, some of this degeneracy can be lifted. The collisional excitation rates for CH$_3$CN used in the LVG models are derived from state-to-state rates, $Q(L,M)$, calculated by \citet{Green1986} for temperatures up to 140 K. Rates for higher temperatures were obtained by extrapolation. This is generally considered to be more reliable than extrapolating to lower temperatures, since the rates tend to vary more smoothly at high temperatures. However, the uncertainties involved are hard to quantify and we therefore adopt the estimated uncertainties given in the original paper, namely somewhere between 50\% and a factor of two.

\begin{table*}
 \caption{Temperatures and methyl cyanide column densities for IRc2 and the northeastern hot zone derived from analysis of the $K$-ladder emission lines.}
 \label{Table:Analysis-Results}
 \centering
 \begin{tabular}{c c lc c lc c lc}
 \hline\hline
 CH$_3$CN & &
 \multicolumn{2}{c}{Rotation Diagram} & \hspace{5mm} & \multicolumn{2}{c}{Population Diagram} & \hspace{5mm} & \multicolumn{2}{c}{LVG Model Fit} \\
 $J'$$\to$$J''$ & &
 \multicolumn{1}{c}{$T_\mathrm{rot}$} & \multicolumn{1}{c}{$N_\mathrm{beam}$} & &
 \multicolumn{1}{c}{$T_\mathrm{rot}$} & \multicolumn{1}{c}{$N_\mathrm{source}$} & &
 \multicolumn{1}{c}{$T_\mathrm{kin}$} & \multicolumn{1}{c}{$N_\mathrm{source}$} \\
 & &
 \multicolumn{1}{c}{(K)} & \multicolumn{1}{c}{(10$^{15}$ cm$^{-2}$)} & &
 \multicolumn{1}{c}{(K)} & \multicolumn{1}{c}{(10$^{16}$ cm$^{-2}$)} & &
 \multicolumn{1}{c}{(K)} & \multicolumn{1}{c}{(10$^{16}$ cm$^{-2}$)} \\
 \hline
 \\[-3.5mm]
 \multicolumn{10}{c}{IRc2 $(+0\arcsec,+0\arcsec)$} \\
 \hline
 \\[-3mm]
 $6_K$--$5_K$     & & 495 & 14  & & 260 & 20  & & 270 & 7.9 \\[2mm]
 $12_K$--$11_K$ & & 415 & 6.8 & & 230 & 6.3 & & 190 & 1.3 \\[2mm]
 $13_K$--$12_K$ & & 329 & 4.9 & & 220 & 6.3 & & 270 & 1.6 \\[2mm]
 $14_K$--$13_K$ & & 442 & 6.1 & & 290 & 6.3 & & 340 & 1.6 \\[1mm]
 \hline
 \\[-3.5mm]
 \multicolumn{10}{c}{Hot Zone Peak} \\
 \hline
 \\[-3mm]
 $6_K$--$5_K$     & & 517 & 5.7 & & 300 & 16  & & 370 & 5.0 \\[2mm]
 $12_K$--$11_K$ & & 415 & 1.4 & & 550 & 6.3 & & 260 & 2.0 \\[2mm]
 $13_K$--$12_K$ & & 402 & 1.3 & & 400 & 5.0 & & 360 & 0.8 \\[2mm]
 $14_K$--$13_K$ & & 476 & 0.4 & & 510 & 5.0 & & 460 & 1.0 \\[1mm]
 \hline
 \\[-3.5mm]
 \multicolumn{10}{c}{Hot Zone Average} \\
 \hline
 \\[-3mm]
 $6_K$--$5_K$     & & 415 & 5.9 & & 220 & 12  & & 270 & 5.4 \\[2mm]
 $12_K$--$11_K$ & & 313 & 1.7 & & 250 & 4.7 & & 200 & 1.1 \\[2mm]
 $13_K$--$12_K$ & & 313 & 1.4 & & 290 & 4.8 & & 240 & 1.0 \\[2mm]
 $14_K$--$13_K$ & & 418 & 1.2 & & 360 & 5.2 & & 350 & 1.0 \\[1mm]
 \hline
 \end{tabular}
 \tablefoot{Hot zone values are quoted for the peak temperature position (``Peak'') and for the average over all positions (``Average'') within the circled region on each map (see Figs.~\ref{Fig:RT_Trot_Ntot_maps}, \ref{Fig:PD_Trot_Nbeam_maps}, and \ref{Fig:LVG_Tkin_Nbeam_maps}).}
\end{table*}

\begin{figure*}
 \resizebox{\hsize}{!}{\includegraphics{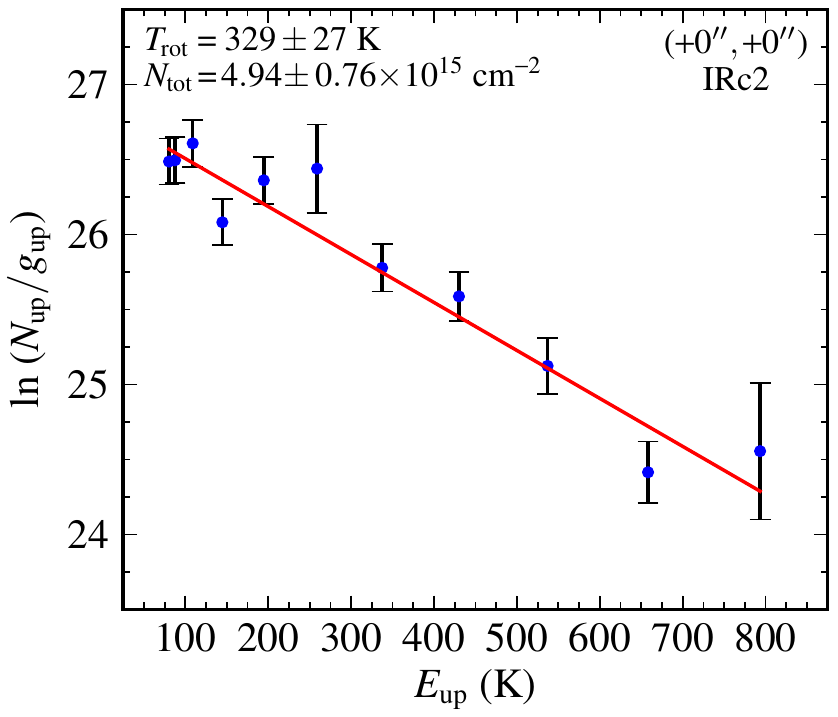}\hspace{5mm}
 \includegraphics{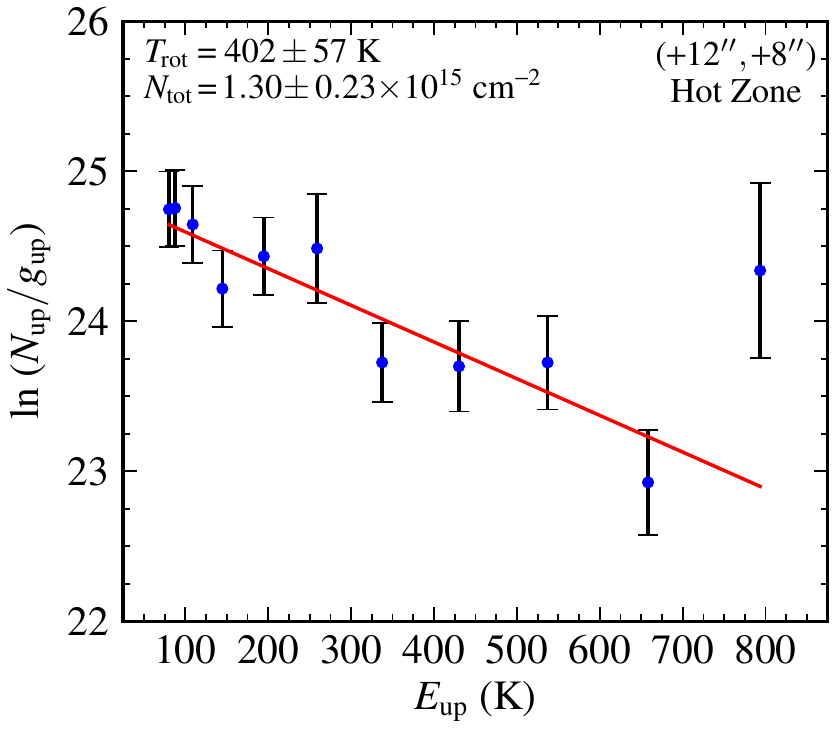}}
 \caption{Rotation diagrams for the CH$_3$CN $J$=$13_K$--$12_K$ lines observed at the centre of the map (\textit{left}), corresponding to the location of IRc2, and an offset position (\textit{right}), corresponding to a temperature peak within the northeastern hot zone.}
 \label{Fig:13-12_rotation-diagrams}
\end{figure*}

All these methods suffer from the additional limitation that they assume constant temperature and abundance in the emitting gas, and that the emission from each transition line arises from the same source size. More complex models introducing temperature and abundance gradients are beyond the scope of this study and in any case suffer other issues of degeneracy, requiring some prior knowledge of the source structure in order to constrain them. Properties derived from the techniques included in this study are therefore considered average values for the gas observed within the telescope beam at each map position. In the case where the gas resides in unresolved clumps, these properties are average values for the ensemble of clumps within the beam.

Given the unknown degree of inaccuracy introduced by failing to account for possible gradients or more complicated source structure, it is difficult to quantify the uncertainties on the derived properties in this analysis. Formal errors determined from fitting slopes to the rotation diagrams and from the $\chi^{2}$ minimisation procedure were generally found to be $\lesssim 30$\% for the rotational and kinetic temperatures and less than a factor of 2 for the column densities. These uncertainties rise for positions near the map edges, where the emission is weaker, but we discard these points from our analysis, as discussed in the next section. Since it is difficult to place a value on the uncertainties introduced by the more general assumption of uniform source properties, we prefer to adopt conservative estimates for the uncertainties, namely 30\% for the derived temperatures and a factor of 2--3 for the column densities.


\begin{figure*}
 \centering
 \resizebox{\hsize}{!}{\includegraphics{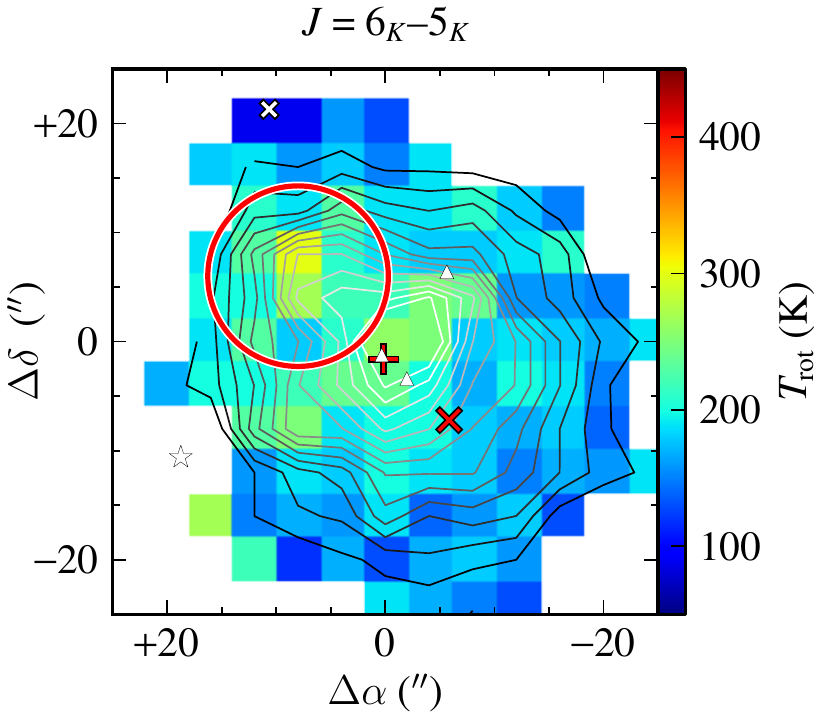}\hspace{5mm}
 \includegraphics{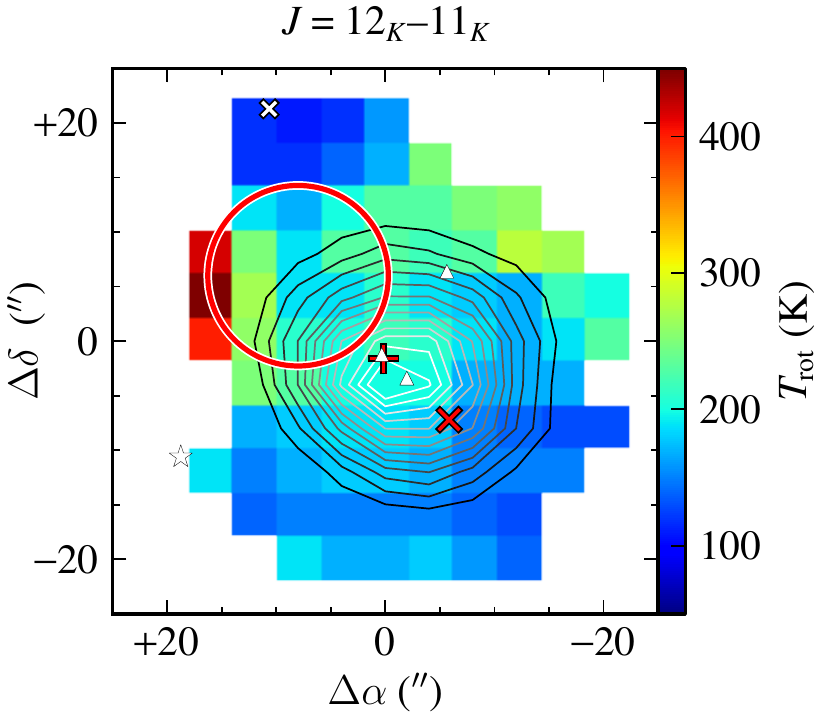}}\\
 \vspace{5mm}
 \resizebox{\hsize}{!}{\includegraphics{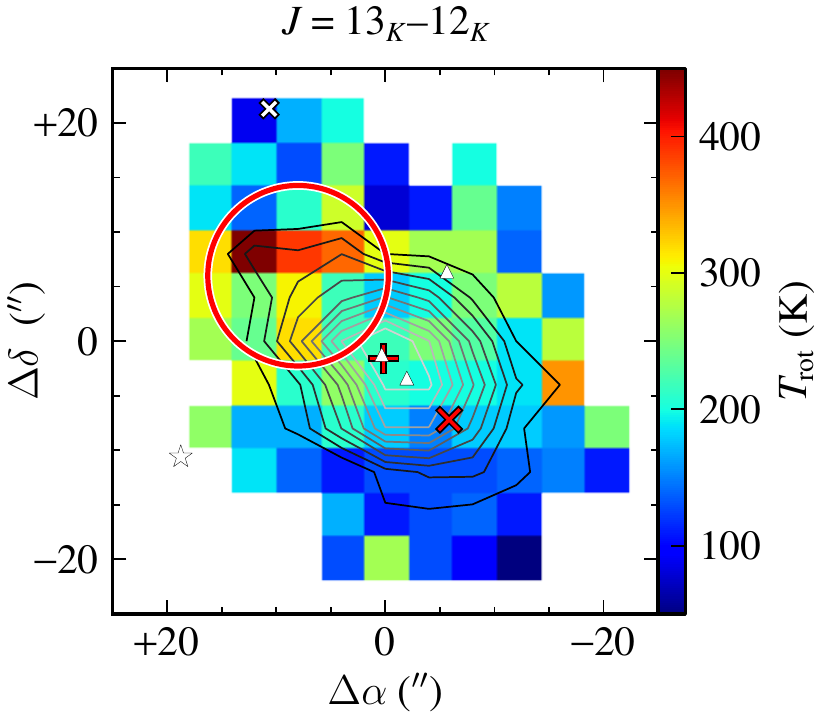}\hspace{5mm}
 \includegraphics{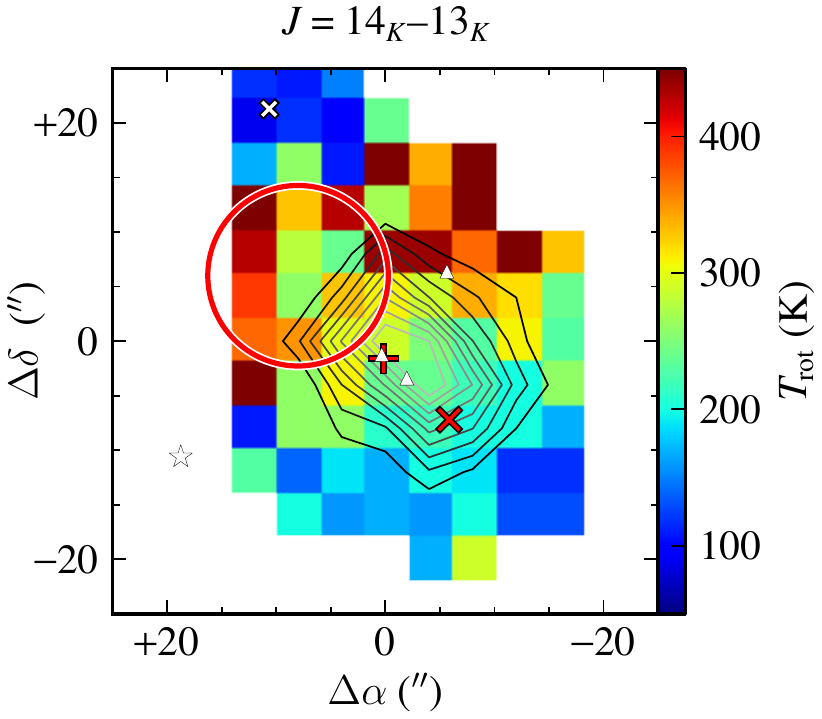}}
 \caption{Results of the population diagram analysis, accounting for the effects of line opacity and beam dilution. Maps of methyl cyanide rotational temperature ($T_\mathrm{rot}$, colour scale) and beam-averaged column density ($N_\mathrm{beam}$, overlaid contours) across the Orion KL region. Both quantities are determined from population diagram analyses of the $J$=$6_K$--$5_K$, $J$=$12_K$--$11_K$, $J$=$13_K$--$12_K$, and $J$=$14_K$--$13_K$ line emission at each map position (see Sect.~\ref{Analysis:Opacity-Dilution} for details). Column density contours are plotted at 0.5$\times$10$^{15}$ cm$^{-2}$ intervals, starting at 1.0$\times$10$^{15}$ cm$^{-2}$ (black lines). The locations of the hot core, compact ridge, radio continuum sources, and the northeastern hot zone are indicated as in previous figures.}
 \label{Fig:PD_Trot_Nbeam_maps}
\end{figure*}

\section{Temperature and column density distributions in Orion KL}\label{Results:Temperature-Maps}

\subsection{Rotation diagram results}\label{Results:Rotation-Diagrams}

Having fitted Gaussians to the methyl cyanide emission lines at each map position (see Sect.~\ref{Analysis:Gaussian-Fits}) and then determining rotational excitation temperatures and total beam-averaged column densities from the resulting rotation diagrams (Sect.~\ref{Analysis:Rotation-Diagrams}), the temperature and column density distributions inferred from the $J$=6--5, 12--11, 13--12, and 14--13 sets of $K$-ladder emission lines are shown in Fig.~\ref{Fig:RT_Trot_Ntot_maps}. In all the maps presented in this section, we have excluded positions where the emission is too weak to obtain reliable fits to the rotation diagrams. To achieve this, we have imposed a lower limit on the integrated intensity of the $K$=5 line in each set of $K$-ladder transitions, only plotting the results for positions where the total $K$=5 integrated intensity determined from the Gaussian fit corresponds to a 3-sigma detection or better (based on the $T_\mathrm{rms}$ and intensity-weighted line width). The choice of the $K$=5 line for this detection limit serves as a reasonable compromise: imposing a stricter detection limit on the higher energy transitions causes too many points to be excluded, while applying a limit to only the lower $K$ lines includes positions where too few lines are detected to be able to reliably determine the gas properties. The resulting temperature and column density distributions should therefore be reliable within the uncertainties discussed in Sect.~\ref{Analysis:Limitations}.

Though the temperature distributions derived from the separate sets of $K$-ladder transitions show some variations, a common feature seen in all the maps is a region of hot ($\sim$300--500 K) gas centred on the hot core (indicated by a red plus sign on each figure) with temperatures dropping off in the direction toward the compact ridge (indicated by a red cross). The warm gas also extends to the northeast of the hot core and in several cases shows a temperature peak at a position approximately 10--15 arcsec away from IRc2, with values reaching above 400~K. The morphology of this extended warm region varies depending on the set of $K$-ladder transitions used to derive the temperatures, but it typically shows an arc-like structure around the northeastern side of IRc2 and an abrupt drop in temperature on its northern-most edge.

The quiescent gas further from the KL nebula is cooler ($< 200$ K), with relatively low column densities ($<$5$\times$10$^{14}$ cm$^{-2}$), but the weak emission in the outer regions of the maps means that the signal-to-noise is significantly worse, and attempts to construct rotation diagrams for some positions lead to unphysical temperatures and column densities. These points have therefore been omitted from the maps, as discussed above. The extended methyl cyanide emission around Orion KL has previously been mapped by \citet{Wilner1994}, who infer a temperature and column density toward the CS1 condensation in the northeastern ridge (marked as a white cross in the figures) of 50--80 K and $\sim$10$^{14}$ cm$^{-2}$, respectively. These values are consistent with those we obtain from the rotation diagram analysis of the various $K$-ladder lines considered here, with temperatures in the range 70--100 K and column densities of 1--2$\times$10$^{14}$ cm$^{-2}$.

The region around the BN source displays significantly higher temperatures than the quiescent material, typically 350--400 K, and column densities of order 10$^{15}$ cm$^{-2}$, intermediate between those inferred for the hot core and those for the extended ridge. Since the separation between IRc2 and BN ($\approx$9$\arcsec$) is less than the beam size of the telescope used to obtain these maps, the temperature and column densities we derive for the BN position may be slightly affected by emission from the hot core that falls within the beam. However, there is clear evidence that the methyl cyanide around BN exhibits raised temperatures and somewhat higher column densities than the surrounding medium, consistent with heating by the embedded BN object.

The region observed toward the central position of the maps is known to host a number of distinct kinematic sources, including emission from the hot core and plateau, amongst others. Since we analyse the total line emission at each map position, rather than deriving properties for individual kinematic components, our resulting temperatures and column densities for this crowded region are average values from all sources within the beam. A detailed analysis of the separate kinematic components seen at the central position has been carried out in a separate study \citep{Crockett2014} and finds that the derived properties do indeed differ significantly for the distinct source components. At positions further from the centre, however, the emission is typically dominated by a single kinematic component.

The area of hot gas to the northeast of IRc2, which we will subsequently refer to as the ``hot zone'', appears in all the temperature maps obtained from the rotation diagram analysis. Peak temperatures in this region vary in location from map to map, but typically occur at or near the position $(+12\arcsec,+8\arcsec)$ relative to the map centre, and always within one beam width of this point. Accounting for the variations in the shape and position of the hot zone feature seen in the temperature maps produced by each analysis method, its approximate extent is indicated by a red circle on Fig.~\ref{Fig:RT_Trot_Ntot_maps} and on subsequent maps.

Rotation diagrams for the methyl cyanide $J$=$13_K$--$12_K$ lines observed toward IRc2 and toward the peak temperature position within the northeastern hot zone are shown in Fig.~\ref{Fig:13-12_rotation-diagrams}. Table~\ref{Table:Analysis-Results} lists the rotational temperatures and beam-averaged CH$_3$CN column densities we derive at the location of IRc2 from the $6_K$--$5_K$ to $14_K$--$13_K$ sets of $K$-ladder lines. The peak temperatures and corresponding beam-averaged CH$_3$CN column densities that we find within the hot zone, in addition to the average values for this region, are also listed. We adopt conservative estimates for the uncertainties of 30\% for the temperature values and a factor of 2--3 for the column densities, as discussed in Sect.~\ref{Analysis:Limitations}.

The upper state column densities for the $K$=10 line on the rotation diagrams in Fig.~\ref{Fig:13-12_rotation-diagrams} (rightmost points) appear to fall outside the general trend displayed by the lower-$K$ lines, especially so for the hot zone position to the northeast of IRc2, where the $K$=10 point is at least a factor of three higher than the $K$=9 value. This deviation may be due to a baseline ripple or other artefact in the raw spectra. If the $K$=10 value is excluded from the fit, then the resulting rotational temperature and beam-averaged column density drop to 361$\pm$65~K and 8.7$\pm$2.7$\times$10$^{14}$~cm$^{-2}$. These values are consistent with those obtained with the $K$=10 point included, within the fit uncertainty, and the temperature remains significantly higher than the surrounding region. For IRc2, removing the $K$=10 point leads to a change of only a few degrees in the rotational temperature and negligible change in the best-fit column density.

\begin{figure}
 \resizebox{\hsize}{!}{\includegraphics{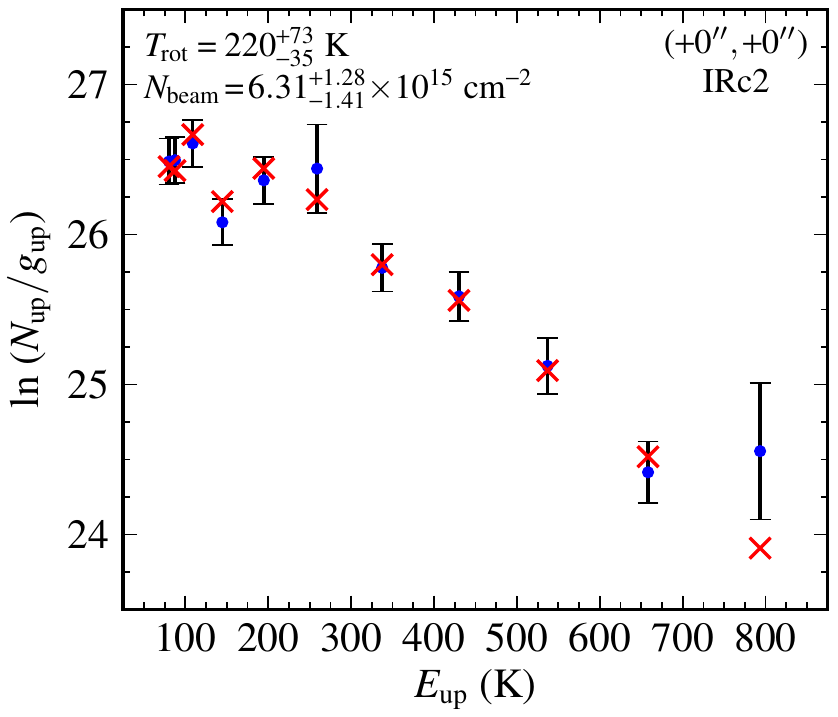}}\\
 \resizebox{\hsize}{!}{\includegraphics{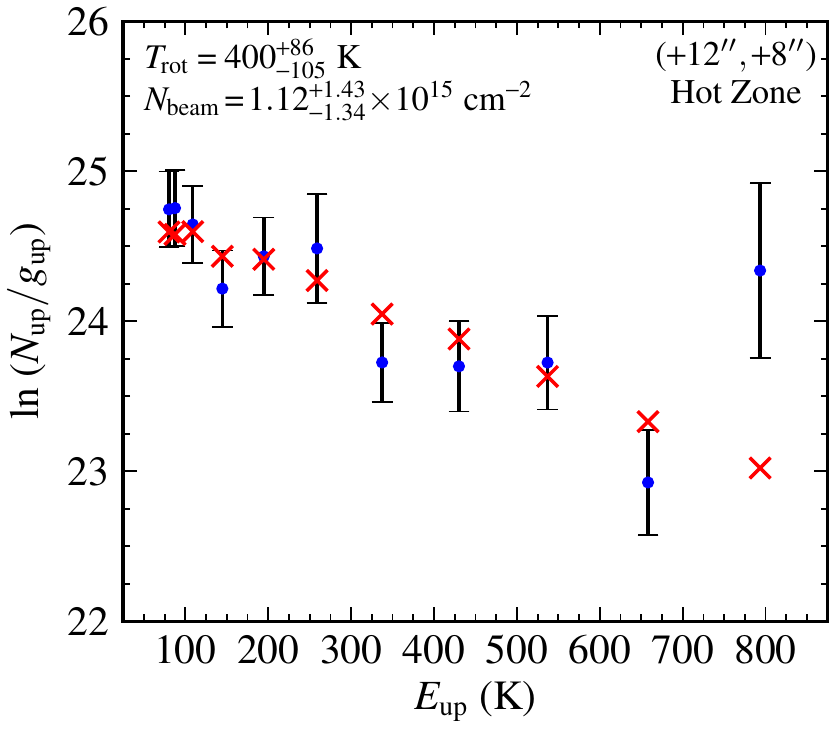}}
 \caption{Population diagrams comparing observations (blue dots) and best-fit LTE model predictions (red crosses) for CH$_3$CN $J$=$13_K$--$12_K$ lines at the map centre (\textit{top}), corresponding to the location of IRc2, and the offset position (\textit{bottom}), corresponding to a temperature peak within the northeastern hot zone.}
 \label{Fig:13-12_population-diagrams}
\end{figure}

The temperatures and column densities derived from the $J$=$6_K$--$5_K$ lines are significantly higher than those derived from the higher $K$-ladder transitions. The higher temperatures are likely due to the fact that the $6_K$--$5_K$ lines have low upper state energies ($<$200 K) and so cannot reliably constrain the temperature when the gas is warm, as is the case in the central region of Orion KL. Conversely, the larger column densities obtained indicate that these lines probe more of the cool gas along the line of sight, owing to their lower critical densities and energies.

\subsection{Population diagram results}\label{Results:Population-Diagrams}

The distributions of rotational excitation temperature and beam-averaged methyl cyanide column density that we obtain from the population diagram analysis of the $J$=6--5, 12--11, 13--12, and 14--13 sets of $K$-ladder emission lines are shown in Fig.~\ref{Fig:PD_Trot_Nbeam_maps}. The results obtained from this analysis account for line opacity and beam dilution effects (see Sect.~\ref{Analysis:Opacity-Dilution}), so represent a more detailed treatment of the emission with respect to the pure rotation diagram technique (Fig.~\ref{Fig:RT_Trot_Ntot_maps}). The temperatures derived in this analysis are typically lower than those obtained from the rotation diagrams discussed above. The main reason for this is the consideration of the line opacity, which can be significant in the $K$=0 and 1 lines, leading to ``flatter'' slopes in the rotation diagrams and correspondingly higher excitation temperatures, as discussed in Sect.~\ref{Analysis:Limitations}. Accounting for the line opacity therefore reduces the inferred excitation temperatures.

We find that the temperature and column density distributions derived from the population diagram analysis are broadly similar to those obtained from rotation diagrams. Most notably, the hot zone again appears as a high temperature region to the northeast of IRc2. Some features become more prominent while others are no longer present, due to the introduction of line opacity into the analysis and the weak degeneracy that exists between opacity and excitation temperature (see Sect.~\ref{Analysis:Limitations}). These differences are unlikely to be due to large uncertainties in the fitting procedure, since positions with weak or undetected lines have been excluded from the maps, as discussed at the start of this section.

\begin{figure*}
 \centering
 \resizebox{\hsize}{!}{\includegraphics{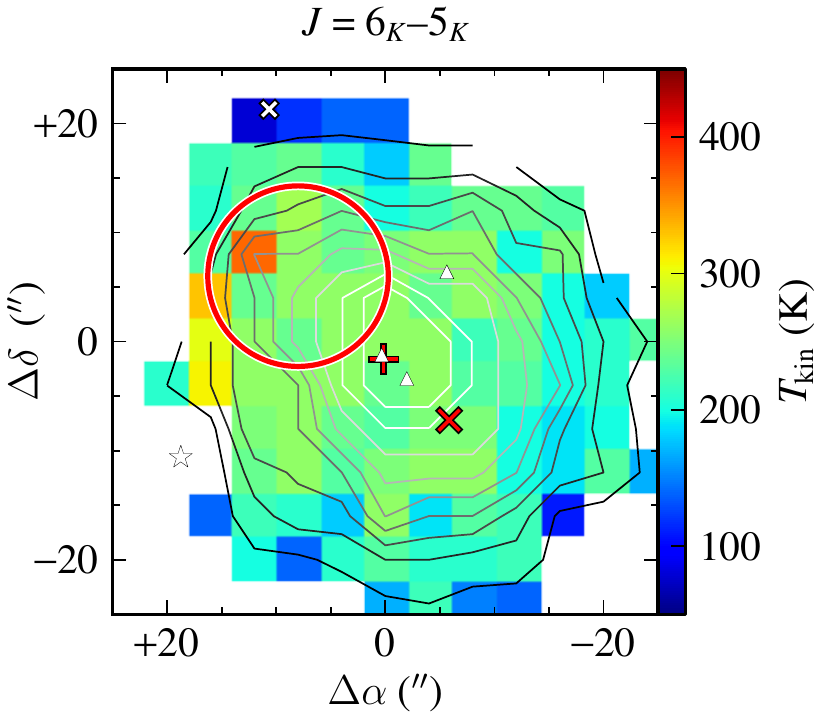}\hspace{5mm}
 \includegraphics{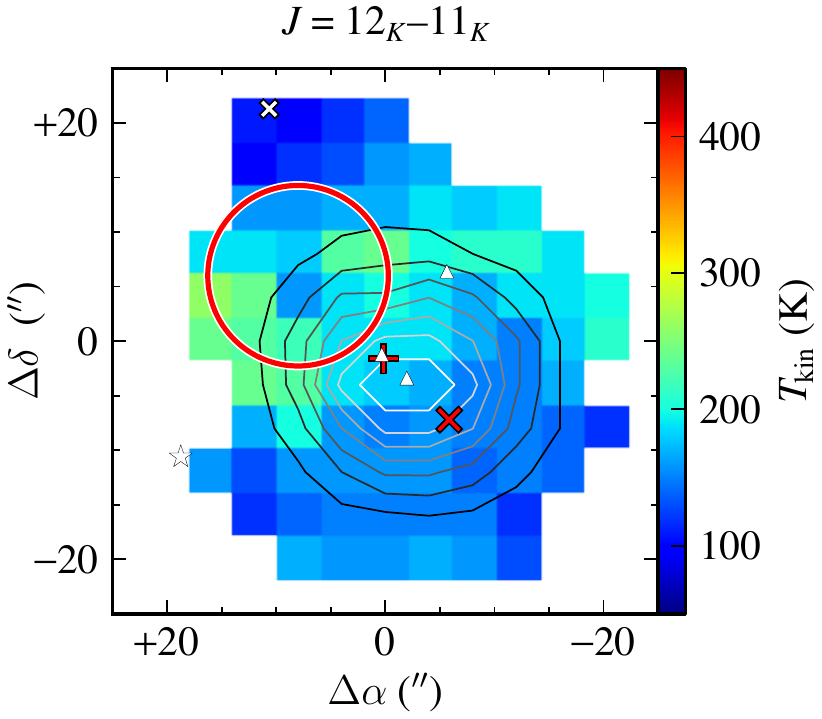}}\\
 \vspace{5mm}
 \resizebox{\hsize}{!}{\includegraphics{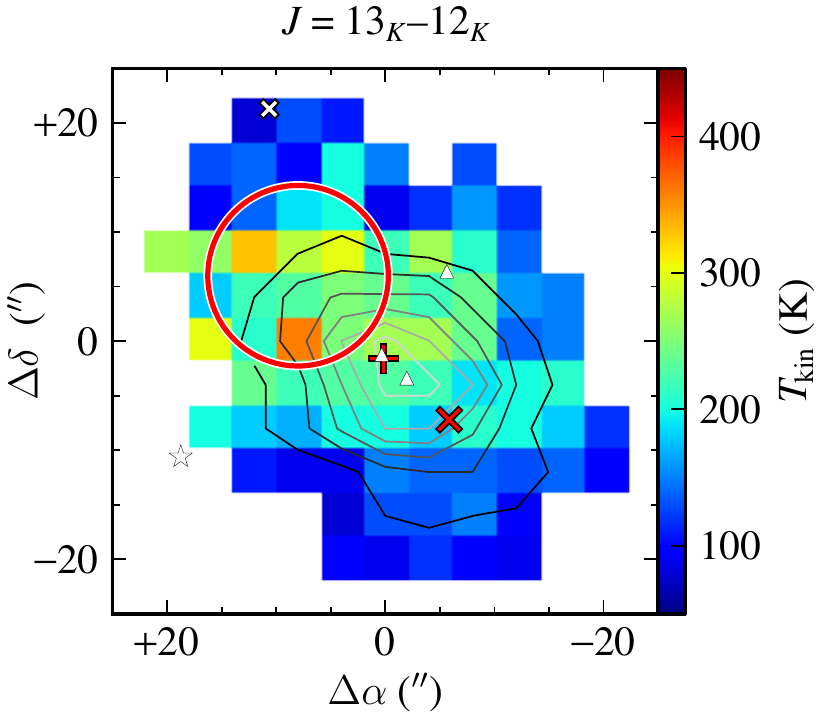}\hspace{5mm}
 \includegraphics{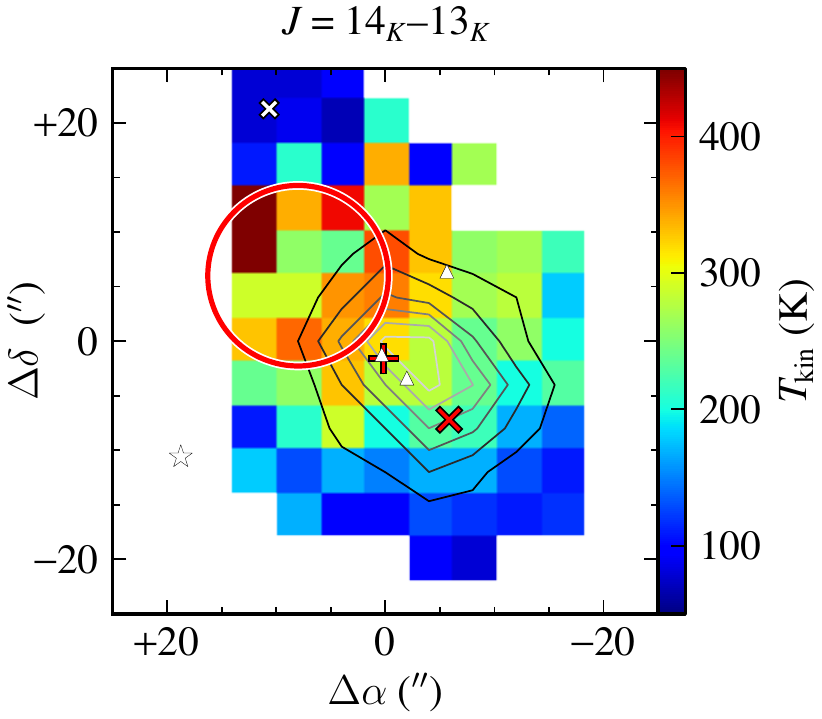}}
 \caption{Results of the LVG model fits. Maps of the kinetic temperature ($T_\mathrm{kin}$, colour scale) and beam-averaged column density ($N_\mathrm{beam}$, overlaid contours) across the Orion KL region determined by fitting LVG models to the observed lines of CH$_3$CN $J$=$6_K$--$5_K$, $J$=$12_K$--$11_K$, $J$=$13_K$--$12_K$, and $J$=$14_K$--$13_K$ at each map position (see Sect.~\ref{Analysis:LVG-Model-Fits} for details). Column density contours are plotted at 0.5$\times$10$^{15}$ cm$^{-2}$ intervals, starting at 0.5$\times$10$^{15}$ cm$^{-2}$ (black lines). The locations of the hot core, compact ridge, radio continuum sources, and the northeastern hot zone are indicated as in previous figures.}
 \label{Fig:LVG_Tkin_Nbeam_maps}
\end{figure*}

Population diagrams for the $J$=$13_K$--$12_K$ lines observed toward IRc2 and the northeastern hot zone are shown in Fig.~\ref{Fig:13-12_population-diagrams}. The uncertainties listed on the diagrams correspond to the 1-sigma confidence limits from the $\chi^{2}$ distributions and give an indication of how well the properties are constrained by the fits; however, the more conservative uncertainty estimates given in Sect.~\ref{Analysis:Limitations} dominate over these formal 1-sigma limits. From the population diagram analysis of each set of $K$-ladder emission lines, we list the rotational temperatures and source-averaged methyl cyanide column densities that we obtain for IRc2 and the hot zone in Table~\ref{Table:Analysis-Results}. The population diagram analysis of the emission toward the hot zone yields rotational temperatures of 200 to 550 K and source-averaged column densities in the range 10$^{16}$ to 10$^{17}$~cm$^{-2}$.

The determination of the beam dilution suffers some degeneracy since we analyse each set of $K$-ladder lines separately (the LVG results discussed in a later section include simultaneous fits to multiple sets of $K$-ladder transitions and are therefore better able to constrain this parameter). That said, we obtain beam dilution factors typically of order 0.1 in the central region around the hot core and compact ridge, corresponding to source sizes of a few arcsec. Further from the centre, the dilution factor drops to $f_\mathrm{beam} \lesssim 0.01$ (source sizes below 1 arcsec). This suggests that the methyl cyanide resides in small unresolved clumps, as has been suggested previously for the hot core \citep{Beuther2005, Wang2010}, and that this is likely also the case for the more extended emission seen in the quiescent cloud.

Line opacities for the $K$=0 transition (the most optically thick line in each $K$-ladder) determined from the population diagram analysis are typically $\gtrsim$1 near the map centre, i.e., in the vicinity of the hot core and compact ridge, forming a region of high opacity emission that extends in a northeast-southwest direction encompassing these two sources. The $J$=$13_K$--$12_K$ lines also suggest that this region of high opacity might extend toward BN, though the other $K$-ladder results do not reflect this. In the vicinity of the hot zone, population diagram analysis yields opacities of $\sim$0.5, indicating optically thinner emission falling just outside the region of high opacity. However, the $J$=$6_K$--$5_K$ line analysis shows increased opacity (closer to 1) within the hot zone itself. The line opacities quickly drop off to $\ll$1 outside of the central region, though emission toward the CS1 condensation in the extended ridge shows higher opacity ($\sim$0.1--0.5).

Since the line opacities are significant in the central part of the maps, the temperatures derived from the population diagram method are likely to be more reliable in this region than those obtained with the rotation diagram technique in the previous section. As discussed above, the inferred rotational temperatures are generally lower when the line opacity is accounted for (see Table~\ref{Table:Analysis-Results}). These temperatures are also consistent with those obtained from LVG models (see next section) and are therefore considered more accurate. Further from the centre, lower column densities mean that similar temperatures are derived both with and without accounting for the line opacity; however, the population diagram analysis of the hot zone emission is likely to yield more reliable estimates for the temperatures in that region, since the opacities are again higher.

\subsection{LVG model results}\label{Results:LVG-Models}

Finally, we show the kinetic temperature and beam-averaged methyl cyanide column density distributions derived from our LVG model analysis in Fig.~\ref{Fig:LVG_Tkin_Nbeam_maps}. The four sets of $K$-ladder transitions are first modelled separately, producing temperature maps that are generally similar to those obtained from the population diagram analysis, in particular the presence of the northeastern hot zone. Column densities peak in the hot core, typically somewhere between the continuum sources \textit{I} and \textit{n}, but a slight elevation toward the hot zone is also seen in some $K$-ladders. In addition, LVG models allow the density of the emitting gas to be constrained. The derived number densities are a few $\times$10$^{6}$ cm$^{-3}$ or higher in both the hot core and compact ridge, falling to $\sim$10$^{5}$ cm$^{-3}$ in the quiescent gas of the extended ridge and lower elsewhere in the region.

\begin{figure}
 \resizebox{\hsize}{!}{\includegraphics{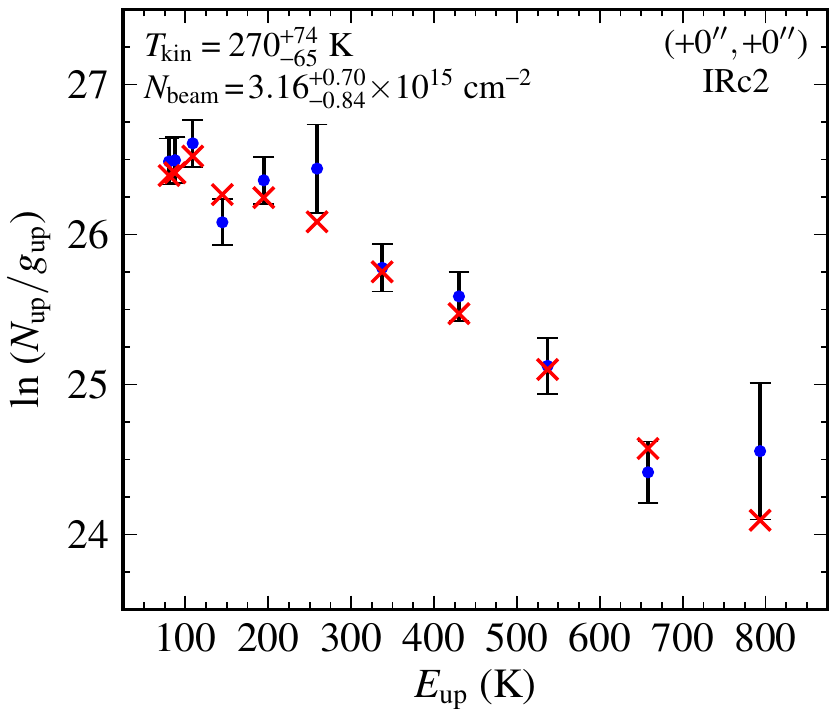}}\\
 \resizebox{\hsize}{!}{\includegraphics{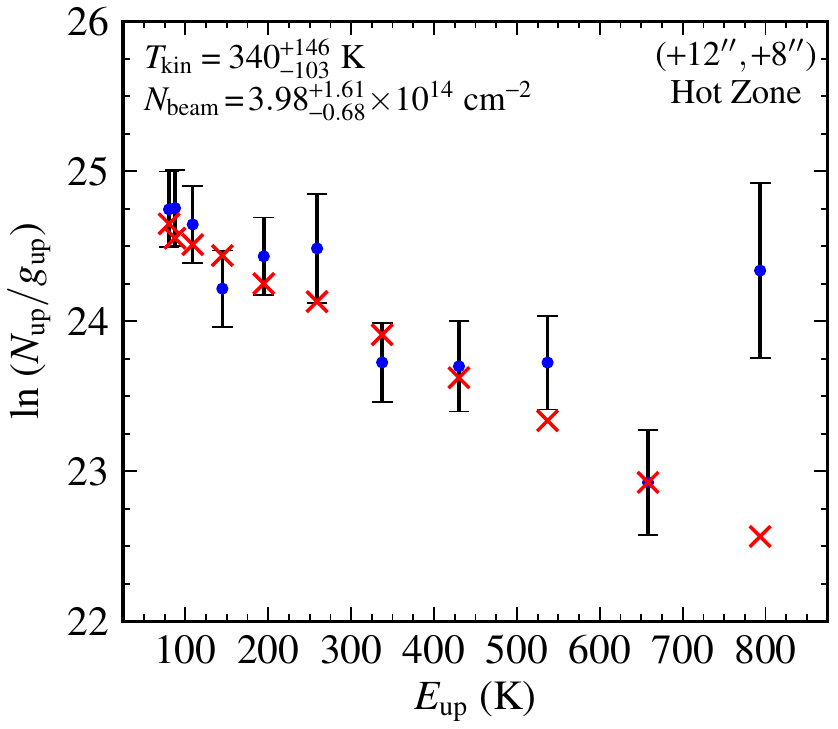}}
 \caption{Population diagrams comparing observations (blue dots) and best-fit LVG model predictions (red crosses) for CH$_3$CN $J$=$13_K$--$12_K$ lines at the map centre (\textit{top}), corresponding to the location of IRc2, and the offset position (\textit{bottom}), corresponding to a temperature peak within the northeastern hot zone.}
 \label{Fig:13-12_LVG-population-diagrams}
\end{figure}

Best-fit model results for the $J$=$13_K$--$12_K$ lines are shown represented as population diagrams in Fig.~\ref{Fig:13-12_LVG-population-diagrams} for the IRc2 and hot zone positions. Inferred kinetic temperatures and source-averaged CH$_3$CN column densities are listed in Table~\ref{Table:Analysis-Results}. We note that excluding the $K$=10 line from the population diagram and LVG model fits for the hot zone position shown in Figs.~\ref{Fig:13-12_population-diagrams} and \ref{Fig:13-12_LVG-population-diagrams} yields temperatures of 310--350~K, which remain consistent with the values listed in Table~\ref{Table:Analysis-Results} (within 1-sigma uncertainties). The population diagram and LVG model fit properties for IRc2 remain unchanged when the $K$=10 point is removed.

The source filling factors (i.e., beam dilution factors) given by the best-fit LVG models indicate the clumpy nature of the CH$_3$CN emitting gas in Orion (see Appendix~\ref{Appendix:Beam-Dilution}). In the region surrounding the hot core and compact ridge, we infer dilution factors of 0.05--0.3, corresponding to source sizes of $\sim$3--6 arcsec. Interferometric observations of methyl cyanide in Orion KL \citep[e.g.,][]{Wang2010} appear to spatially resolve the emission with an arcsecond beam, however, the derived source filling factors from those observations were found to be smaller than unity, suggesting that the observed structures may actually be ensembles of even smaller clumps. Alternatively, such low filling factors could be the result of missing flux. Given the larger beam size of our single-dish observations, the source filling factors we obtain may therefore be upper limits, more representative of the arcsecond-sized structures.

Toward the hot zone, the best-fit LVG models have source filling factors of 0.025--0.05, corresponding to source sizes of $\sim$1.5--3 arcsec. The beam dilution maps generally show a region of low filling factor ($<$0.05) around the hot zone. This may imply that the emitting gas in this region resides in similarly sized clumps to those found in the centre, but more sparsely distributed, or alternatively, that the clumps in this region are about half the size of those seen in the hot core and compact ridge. Toward the extended ridge and the CS1 clump within it, we find source filling factors of 0.04--0.15, giving source sizes of $\sim$2--5 arcsec, similar to the size found by \citet{Wilner1994} for CS1. Outside of the central region and extended ridge, the filling factor drops off rapidly to 0.01 or below, though these values are rather unconstrained since the line opacity is low.

\subsection{Combined LVG model results}\label{Results:Combined-Models}

\begin{figure}
 \resizebox{\hsize}{!}{\includegraphics{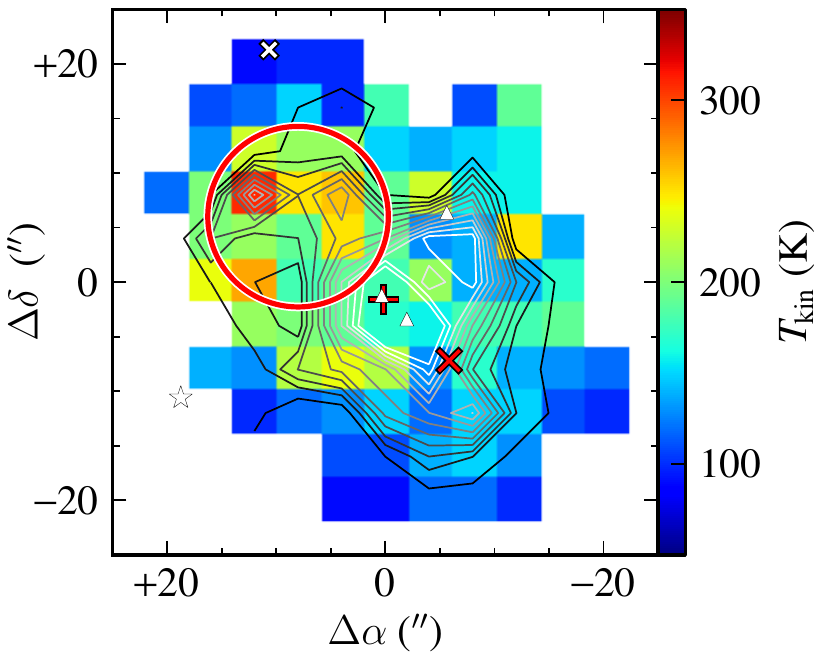}}\\
 \resizebox{\hsize}{!}{\includegraphics{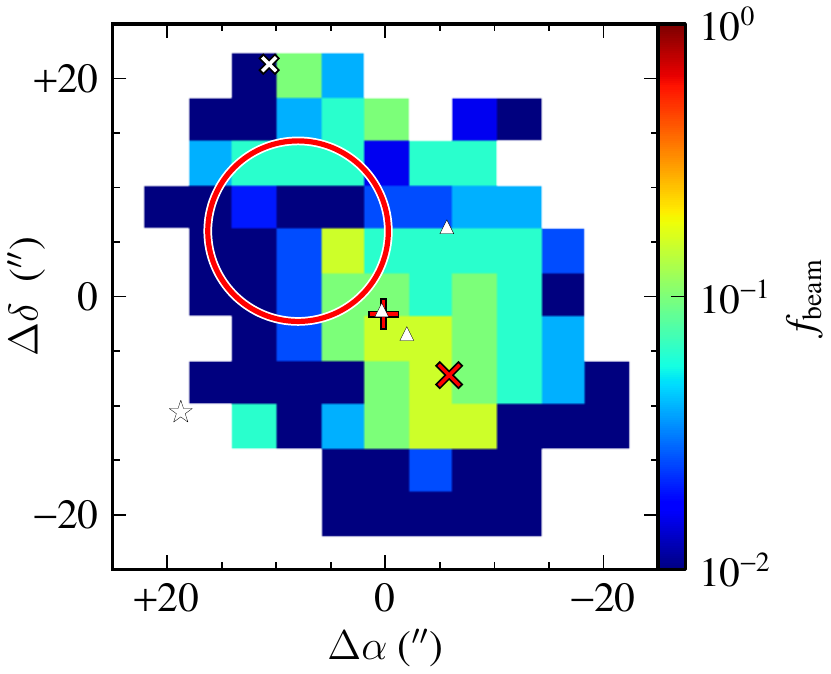}}\\
 \resizebox{\hsize}{!}{\includegraphics{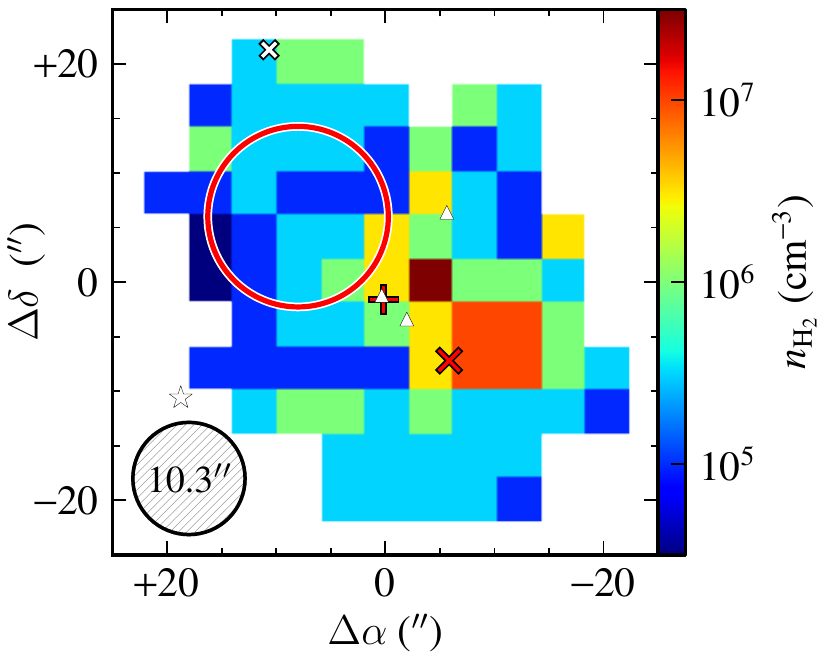}}
 \caption{Maps of the kinetic temperature ($T_\mathrm{kin}$, \textit{top}), beam-averaged methyl cyanide column density ($N_\mathrm{beam}$, contours), beam dilution factor ($f_\mathrm{beam}$, \textit{middle}), and H$_2$ number density ($n_\mathrm{H_2}$, \textit{bottom}) across the Orion KL region, determined by simultaneously fitting LVG models to the observed $J$=$12_K$--$11_K$, $J$=$13_K$--$12_K$, and $J$=$14_K$--$13_K$ lines at each position. Column density contours are plotted at 0.5$\times$10$^{15}$ cm$^{-2}$ intervals, starting at 1.0$\times$10$^{15}$ cm$^{-2}$ (black). The locations of the hot core, compact ridge, continuum sources, and hot zone are indicated as in previous figures.}
 \label{Fig:LVG_combo_maps}
\end{figure}

In order to better constrain the beam dilution and volume densities, the data for the $J$=$12_K$--$11_K$, $13_K$--$12_K$, and $14_K$--$13_K$ lines were fitted simultaneously with a single LVG model at each map position. The $J$=$6_K$--$5_K$ lines were not included in the fit since doing so led to unrealistically low kinetic temperatures ($\sim$100 K for IRc2), which we attribute to the lower energy $6_K$--$5_K$ transitions dominating the goodness of fit and biasing the temperature downward. Since the angular resolution of the $J$=$6_K$--$5_K$ maps is significantly less than that of the higher frequency transitions, there is little additional information lost by excluding these lines from the fit.

The resulting maps of kinetic temperature, beam-averaged column density, beam dilution factor, and H$_2$ number density derived from the simultaneous LVG model fits to the $J$=$12_K$--$11_K$, $13_K$--$12_K$, and $14_K$--$13_K$ lines at each map position are shown in Fig.~\ref{Fig:LVG_combo_maps}. These maps clearly reveal the presence of the northeastern hot zone (with $T_\mathrm{kin} \!\approx\! 300$~K) and indicate kinetic temperatures for IRc2 and the hot core of 170--230~K. The methyl cyanide column density peaks just south of IRc2, close to source \textit{n}, at $\sim$8$\times$10$^{16}$~cm$^{-2}$ (corrected for beam dilution). Another column density peak appears to the south of the BN object, though the gas is cooler here, at 130 K. The column density also rises again in the hot zone region and coincides with the temperature peak.

The source filling factor distribution is largely the same as that derived from the LVG model fits to the individual $K$-ladders. An overall picture of clumpy structure on large scales therefore emerges from the various analyses of these methyl cyanide maps, consistent with other studies of Orion KL. We note, however, that the derived filling factors are rather poorly constrained in some regions. In the central part of the map where we find high line opacities, the uncertainty on the source filling factor is typically less than a factor of 2, but further from the centre the lines are optically thinner and in areas outside of the ridge the filling factor is essentially unconstrained, having uncertainties of an order of magnitude or more.

The temperature and column density distributions within the area of the northeastern hot zone reveal several distinct peaks. Similar peaks were also seen in this region for some of the distributions obtained from the previous analyses. Given this varying morphology, it is unlikely that a single, well-defined source is responsible for the generally high temperatures seen in the hot zone. Instead, the inferred distributions are better explained by several unresolved hot clumps situated within a common warm environment. The slight variation in the positions of the temperature peaks are therefore the result of the limited spatial resolution of the maps and uncertainties in the methodology applied.

Finally, we show the H$_2$ number density distribution derived from the simultaneous LVG model fits in the bottom panel of Fig.~\ref{Fig:LVG_combo_maps}. High densities ($\ge$10$^{7}$~cm$^{-3}$) are found in the central region close to the hot core and compact ridge, though they seem to peak slightly to the west of both sources. Intermediate densities of order 10$^{6}$~cm$^{-3}$ are present in the area surrounding these two sources and trace the general shape of the extended ridge to the northeast, rising again slightly near the CS1 condensation. In and around the hot zone, the number densities are somewhat lower, at a few~$\times\,$10$^{5}$~cm$^{-3}$, and are similar to those found in the outer regions of the maps, where densities of about 10$^{5}$~cm$^{-3}$ are maintained across most of the cloud. The inferred densities generally have uncertainties of order $\pm$0.5 dex, based on the 1-sigma confidence limits from the $\chi^{2}$ distributions.


\subsection{Abundance distribution of methyl cyanide}\label{Results:Abundance-Map}

In order to compare the methyl cyanide emission maps with chemical model predictions, we have derived the CH$_3$CN abundance distribution from the best-fit LVG model results for the simultaneous fit to the three sets of $K$-ladder lines. To do so, we have used a SHARC 350\,$\mu$m continuum emission map of Orion obtained by \citet[][kindly provided by the authors]{Lis1998} and calculated total H$_2$ column densities, $N(\mathrm{H_2})$, at each position by assuming a dust temperature of 50~K, a dust mass opacity coefficient at 350\,$\mu$m of 10.1~cm$^2$\,g$^{-1}$ \citep[column 6 in Table~1 of][corresponding to agglomerated dust grains with thin ice mantles at densities $\sim$10$^{6}$~cm$^{-3}$]{Ossenkopf1994}, a gas-to-dust mass ratio of 100, and that the gas is fully molecular. The beam-averaged CH$_3$CN column density map shown in Fig.~\ref{Fig:LVG_combo_maps} was then divided by the $N(\mathrm{H_2})$ map to obtain the abundance distribution across the Orion KL nebula. The SHARC continuum data have an effective angular resolution of 12$\arcsec$, with 4$\arcsec$ spacing between map points, making them ideally suited for comparison with our methyl cyanide maps. Due to the similar beam sizes, we have used beam-averaged column densities to derive the abundances. Adjusting for the beam dilution factors determined from the LVG models, and assuming that the H$_2$ column densities are undiluted in the same beam, the source-averaged abundances would therefore be about an order of magnitude higher.

We note that dust temperatures are likely to be higher in the central region of Orion KL, particularly in the hot core and around neighbouring embedded sources. The densities in this region are sufficiently high that the gas and dust temperatures are coupled. Adopting a dust temperature of 200~K instead for the central part of the map would yield CH$_3$CN abundances about a factor of 5 higher. For the bulk of the extended ridge, including the hot zone, lower dust temperatures are more appropriate and our choice of a constant value of 50 K serves as a reasonable average \citep[e.g.,][]{Lis1998, Dupac2001}.

\begin{figure}
 \resizebox{\hsize}{!}{\includegraphics{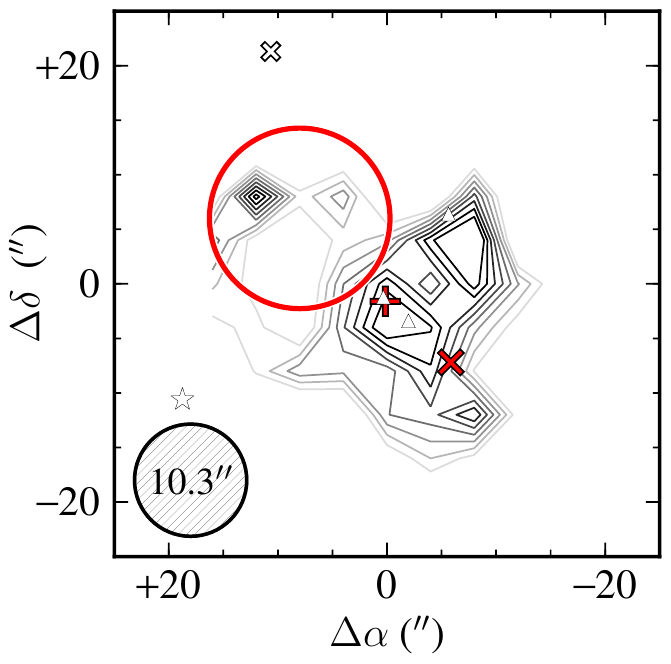}}
 \caption{Contour map of the CH$_3$CN abundance (with respect to H$_2$) across the Orion KL region, derived from beam-averaged methyl cyanide column densities obtained from simultaneous LVG model fits to multiple $K$-ladder transitions (see Sect.~\ref{Results:Temperature-Maps} for details). Contours are plotted from 2$\times$10$^{-9}$ (white) to 6$\times$10$^{-9}$ (black) at 0.5$\times$10$^{-9}$ intervals. The locations of the hot core, compact ridge, radio continuum sources, and the hot zone are indicated as in previous figures.}
 \label{Fig:abundance_map}
\end{figure}

The resulting distribution of methyl cyanide abundance is shown in Fig.~\ref{Fig:abundance_map}. We find abundances of $\sim$6$\times$10$^{-9}$ in the hot core and the northern part of the compact ridge, and a peak value of 7.4$\times$10$^{-9}$ at a position just south of the BN object. The CH$_3$CN abundances across the rest of the cloud are generally much below 10$^{-9}$, with the exception of the peak position of the hot zone and an arc extending southeast of it, in which the abundance rises to 3--6$\times$10$^{-9}$. Toward the location of CS1, the CH$_3$CN abundance is $\sim$10$^{-10}$, consistent with the value found by \citet{Wilner1994}. The fractional abundance we derive for the hot core is about an order of magnitude lower than that determined by \citet{Wang2010} from interferometric observations. However, accounting for beam dilution and higher dust temperatures, the source-averaged abundance is between 10$^{-8}$ and 10$^{-7}$, which is consistent with their results. Similarly, the source-averaged abundance at the peak position within the hot zone becomes $\approx$5$\times$10$^{-8}$, putting it close to the abundances of methyl cyanide produced by C-shock models, as discussed in the following section.


\section{Discussion}\label{Discussion}

\subsection{Cause of the northeastern hot zone}\label{Discussion:Hot-Zone}

The region of hot gas that we detect in all four series of observed $K$-ladder transitions and that we have termed the northeastern hot zone lies roughly midway between IRc2 and a condensation known as CS1 approximately 20--30 arcsec to the northeast of IRc2. This condensation has been observed in both continuum \citep{Mundy1986} and line emission maps \citep{Blake1987, Mangum1993, Wilner1994, Wright1996} and is believed to be a dense clump residing within the ridge, since it displays the same kinematic signature. Interferometric maps of line emission from a number of molecular species indicate that the dense material in CS1 may extend over a region of 10--20 arcsec, with some variation from species to species \citep{Wilner1994, Wright1996, Wiseman1998}. We therefore propose that the hot gas that we infer at a position approximately 15 arcsec to the northeast of IRc2 is the result of a shock front as the outflowing gas originating in the vicinity of IRc2 impacts the southwestern edge of the CS1 condensation. The heating due to this passing shock front is thus responsible for the high temperatures we find in this region.

This scenario is supported by the general morphology and location of the hot zone. The low-velocity outflow in Orion KL is traced by emission from species such as SiO \citep{Plambeck2009}, which shows a clear bipolar morphology in the northeast-southwest direction centred on source \textit{I}. The size of this region is seen to vary in different species, but generally extends about 10--20 arcsec to the northeast of IRc2, consistent with what we find for methyl cyanide. Similarly, \citet{Pardo2005} found that [C\,\textsc{i}] 690\,$\mu$m emission traces an arc around the northeast edge of IRc2 and the hot core (see their figure 2), and also associated this emission with the outflow component. The shape of this emission largely agrees with that of the hot methyl cyanide seen in our maps, showing a bright region to the northeast of IRc2 that begins at approximately the same location as the hot zone and extends toward the CS1 condensation. The clumpy and extended nature of the warm gas we find within the hot zone is also consistent with outflowing material impacting various unresolved clumps in the quiescent ridge. There is therefore good evidence to suggest that this hot zone is linked to the low-velocity outflow.

Emission features in the vicinity of the hot zone have been detected previously by several authors. \citet{Habing1991} proposed that the peak that they detected $\sim$10$\arcsec$ northeast of the hot core in the $J$=$19_K$--$18_K$ lines of methyl cyanide was a previously undiscovered embedded source, likely a young protostar. However, \citet{Wilner1994} argued that it was more likely that this emission peak is produced by an interaction zone in which the dense quiescent gas of the ridge is being impacted by the low-velocity outflow from Orion KL. This second interpretation is in line with our proposed shock heating scenario and seems more likely, since the properties we derive for the gas in the hot zone (broad line widths, high temperatures, relatively low column densities, and clumpy structure) are more consistent with the presence of shocked gas than an embedded source. Indeed, studies based on the dust and line emission from other molecules have also lead to similar conclusions \citep[e.g.,][]{Masson1988, Wright1992}.

\begin{figure*}
 \centering
 \includegraphics[width=17cm]{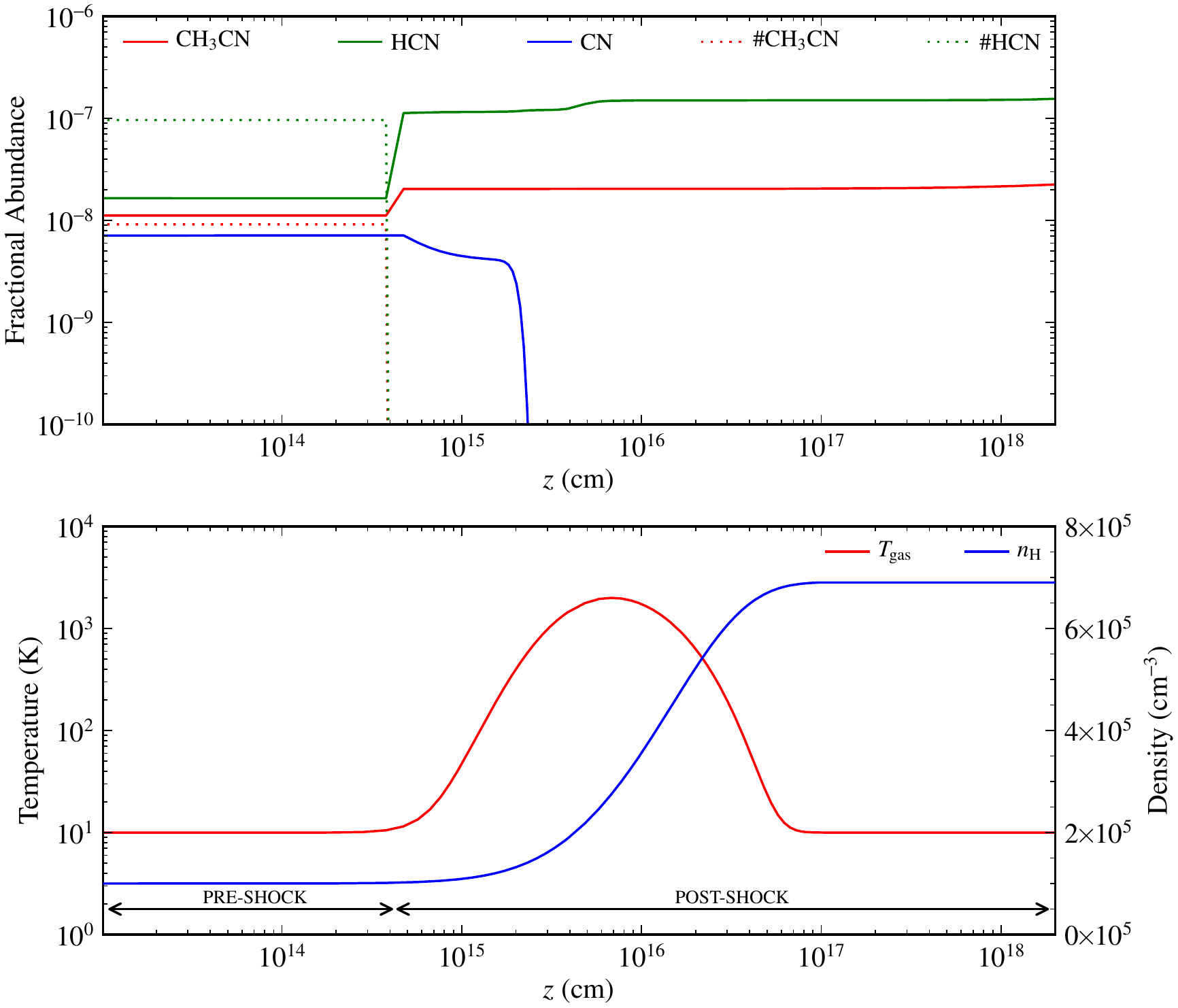}
 \caption{Results from a C-shock chemical model with shock velocity $v_\mathrm{s}=30$ km\,s$^{-1}$ and pre-shock gas-phase depletion of 60\% (see Sect.~\ref{Discussion:Formation} for details). \textit{Top}: Fractional abundances of gas-phase CH$_3$CN, HCN, and CN, and grain surface CH$_3$CN and HCN (indicated by the \# prefix) as a function of distance into the shock front. \textit{Bottom}: Profiles of gas temperature (red) and total hydrogen number density (blue) as a function of distance into the shock front.}
 \label{Fig:Shock-Models}
\end{figure*}

Furthermore, there have been various studies suggesting an association between methyl cyanide emission and outflowing gas in other sources, such as \citet{Codella2009}, \citet{Leurini2011}, and \citet{Moscadelli2013}. \citet{Csengeri2011} also propose that the weak and extended CH$_3$CN emission observed in DR21(OH) is tracing warm gas associated with low-velocity shocks. This further supports the association of the hot methyl cyanide northeast of IRc2 with shocked gas.

The proposed role of the prior explosive event in heating and shaping the KL region \citep{Zapata2009, Zapata2011, Moeckel2012} could in theory be responsible for creating the hot zone, if ejected material from the explosion were colliding with the extended ridge at that location. However, the kinematics of the methyl cyanide emission at this position tie it more convincingly to the low-velocity outflow originating from source \textit{I}.

We can also rule out infrared pumping as a possible explanation for the elevated temperatures in the hot zone. In this scenario, the strong infrared continuum emitted by IRc2 and neighbouring sources could excite the high energy states of the molecule. However, emission from the $\nu_8$=1 vibrationally excited state of methyl cyanide, which is known to be excited by the strong IR continuum within the hot core, is entirely absent in the vicinity of the hot zone. Other mechanisms that might lead to super-thermally excited emission, such as non-Maxwellian state-to-state chemistry as discussed by \citet{Godard2013} for the case of CH$^{+}$, cannot be currently tested, since the gas-phase formation routes of methyl cyanide are still rather uncertain. Based on the location of the hot zone, the broad line widths and high temperatures we derive in that region, and its similar morphology to other tracers of the low-velocity outflow from IRc2, we argue that shock heating caused by the impact of the low-velocity outflow with the edge of the CS1 condensation is the most plausible scenario.


\subsection{Formation mechanisms for methyl cyanide}\label{Discussion:Formation}

The main formation route of methyl cyanide is still uncertain. It is generally believed to be the product of grain surface chemistry, primarily through the reaction of CH$_3$ and CN in grain mantles, followed by its evaporation at temperatures above $\sim$90 K \citep{Garrod2008a}, though modest amounts of CH$_3$CN may also form in the gas phase as HCN reacts with CH$_3^+$.

If the higher temperatures found within the hot zone are indeed caused by shock heating of the gas, we might reasonably ask if methyl cyanide is expected to survive in such environments, or even if its abundance might be enhanced. It has indeed been suggested that methyl cyanide may be enhanced in C-shocks \citep{Codella2009}, where the magnetic precursor to the shock front causes efficient sputtering of dust grains and lower velocities prevent dissociation fronts from forming. With this in mind, we have run a series of chemical models appropriate for C-shocks to explore the abundances of CH$_3$CN that might be produced for a range of parameter space appropriate for this region. The model we adopt is the time-dependent gas-grain chemical model of \citet{Viti2011} that includes an analytical treatment of the physical properties of a passing C-shock and a full treatment of the grain-sputtering and pre- and post-shock chemistry within the region, both in the gas and on the grain surfaces.

The model begins by following the collapse of a diffuse cloud until it reaches a specified pre-shock density, computing the chemistry as gas-phase species freeze out onto grain mantles and are further processed by grain-surface reactions. Grain-surface formation of CH$_3$CN is included in the model through the reaction of CH$_3$ and CN and via hydrogenation of C$_2$N and C$_2$NH$^+$. The degree of depletion of species onto the grain mantles is estimated by calculating the rate per unit volume at which each species freezes out and this is a function of density, temperature, grain parameters, and, of course, the sticking coefficient. Since the latter, as well as the dust size distribution, are unknown, one can control the total freeze-out by a depletion coefficient (which will depend on the abundance of very small grains as well as the binding energy of each species). Non-thermal desorption due to several mechanisms \citep[see][]{Roberts2007a} is also included in the model. The final percentages of freeze-out for our models are 30, 45, and 60\%. Note that the advantage of this approach is that the ice composition is not assumed but is derived by a time-dependent computation of the chemical evolution of the gas-dust interaction process.

After this collapse phase, the passing of a C-shock through the cloud is modelled, with the physical properties of the shocked gas described by a parametric model, taken from \citet{Jimenez-Serra2008}, that depends on the assumed pre-shock gas density and shock velocity $v_\mathrm{s}$. The process of grain mantle sputtering in the shock front is treated in detail by the model, leading to efficient release of the grain mantle species back into the gas phase, and the resulting gas-phase chemistry is followed as the cloud cools after the shock front has passed (see \citealt{Viti2011} for further details). Depletion onto grains from the post-shock gas is not taken into account in the model, but would become significant long after the shock front has passed and the gas has cooled sufficiently. In the hot gas immediately after the passing of the shock front, however, it is safe to assume that depletion is negligible.

To study the possible enhancement of methyl cyanide abundances as a result of low-velocity C-shocks that may be present in the hot zone, we consider a number of models in which we vary the degree of depletion onto grains and the velocity of the passing shock. We assume a pre-shock gas density of 10$^{5}$ cm$^{-3}$ in all the models, shock velocities of 20, 30, and 40 km\,s$^{-1}$, and depletion levels of 30, 45, and 60\%. Figure~\ref{Fig:Shock-Models} shows the resulting fractional abundances for methyl cyanide and associated species for a model with $v_\mathrm{s}=30$ km\,s$^{-1}$ and 60\% depletion, in addition to the gas temperature and density profiles assumed in the parametric shock model. The shock front appears at $\sim$4$\times$10$^{14}$ cm into the model cloud, at which point the methyl cyanide contained in grain mantles (whose abundance is indicated by the label \#CH$_3$CN in the figure) is released into the gas phase due to sputtering of the grain mantles. This produces a twofold increase in the gas-phase abundance of CH$_3$CN, which then remains constant at $\sim$2$\times$10$^{-8}$ throughout the post-shock gas. This jump in methyl cyanide abundance is not dramatic, especially when compared to that of HCN, which increases by an order of magnitude in the shock. The reason for this is that the pre-shock gas-phase abundance of CH$_3$CN is already greater than its grain-surface abundance, due to the relatively low pre-shock density assumed in the model, so even returning the entire mantle reservoir to the gas phase cannot produce a large abundance increase. The degree of freeze-out onto grain mantles depends on a number of model parameters. More CH$_3$CN would be depleted onto grains if either the density were higher, freeze-out were more efficient, or desorption processes were less effective. This would result in a more significant jump in gas-phase methyl cyanide abundance in the post-shock gas.

Comparable methyl cyanide abundances of order 10$^{-8}$ have been inferred previously for the hot core by \citet{Wilner1994} and \citet{Wang2010}. The relatively high abundances attained in our chemical models therefore suggest that high densities and warm environments, such as those found around protostars, are not necessarily the only scenarios that can produce significant quantities of methyl cyanide. Given that the beam-averaged column densities and fractional abundances that we infer for the hot zone in the outflow from IRc2 are higher than those in neighbouring regions, moderate abundance enhancement due to the presence of C-shocks seems plausible.

We find little variation of the post-shock methyl cyanide abundance with shock velocity $v_\mathrm{s}$, since all CH$_3$CN residing on the grain mantles is released back into the gas phase at all velocities considered, so the gas-phase abundance effectively saturates, with no dominant gas-phase reactions that form or destroy it afterwards. Changing the degree of pre-shock depletion of species onto the grains produces a minor change in the post-shock gas-phase abundance, but the limited freeze-out of methyl cyanide in the pre-shock phase due to the relatively low density assumed in the model prevents significant changes, as discussed above. Crucially, the methyl cyanide that is released from grain mantles in the passing shock front remains abundant in the warm post-shock gas and is not rapidly destroyed. C-shock models are thus able to reproduce the methyl cyanide abundances and temperatures that we derive from the emission toward the hot zone. We therefore propose that a shock formation scenario is the most plausible explanation for this component.


\section{Summary and conclusions}\label{Conclusions}

We have obtained full spatially sampled maps of methyl cyanide emission in four sets of its $K$-ladder transitions, $J$=$6_K$--$5_K$, $J$=$12_K$--$11_K$, $J$=$13_K$--$12_K$, and $J$=$14_K$--$13_K$, across the Orion KL region using the IRAM 30m telescope. These maps beautifully display the extended emission from this molecule over an area approximately 40 arcsec across, well beyond the confines of the hot core and compact ridge components. By fitting Gaussian profiles to the line shapes of each set of $K$-ladder transitions, the kinematics of the gas can be studied across the region and reveal a sharp velocity gradient and the distinct kinematic features of the hot core and plateau components. The spectroscopic properties of its emission lines make methyl cyanide an ideal diagnostic of the gas temperature in warm ($>$100 K) and dense ($>$10$^{5}$ cm$^{-3}$) environments. Taking advantage of this, we have used population diagram analysis and LVG modelling to infer the physical conditions [$T_\mathrm{kin}$, $N_\mathrm{tot}$, $n(\mathrm{H_2})$] at each position in our maps. The resulting temperature distributions for the emitting region show a distinct ``hot zone'' surrounding the northeastern edge of the hot core, with inferred temperatures of 300--450 K. This feature is seen in all four sets of $K$-ladder emission lines and emerges from both population diagram and LVG model analyses. The location of the hot zone suggests an association with the northeast-southwest bipolar outflow originating in the vicinity of IRc2. Based on the kinematics and derived physical properties of the methyl cyanide emission in this region, we argue that it is associated with hot shocked gas where the outflow impacts the quiescent material in the extended ridge cloud. We have computed a series of shock chemistry models in order to test this scenario. We find that methyl cyanide can indeed remain present in shock fronts, at high temperatures and elevated abundances. Our main findings are as follows:

\begin{itemize}
\item Methyl cyanide emission is extended across a 40 arcsec region around IRc2, with temperatures of 100--150 K and inferred densities of order $10^{5}$ cm$^{-3}$ in the quiescent gas.
\item The quiescent gas along the molecular ridge shows a global north-south velocity gradient and narrow line widths ($\lesssim$4~km\,s$^{-1}$), while the hot core region at the centre of the maps is clearly identifiable in the CH$_3$CN emission via its distinct kinematic signature, characterised by blue-shifted and broader lines.
\item CH$_3$CN emission peaks at the hot core position; column densities derived from population diagrams and LVG modelling both show that methyl cyanide is over an order of magnitude more abundant in the hot core than in the ambient medium.
\item Temperatures between 200 and 350 K have been determined for the methyl cyanide emission in the hot core and its environs, however, higher temperatures ($\ge$300 K) are seen in a hot zone to the northeast of the hot core.
\item We attribute this feature to shock heating by outflowing gas from active star formation in the vicinity of IRc2. This is supported by the coincidence of this region with emission from shock tracers such as SiO and neutral carbon.
\item Chemical models of C-shocks indicate that methyl cyanide abundances may be enhanced in shock fronts under physical conditions appropriate for this region.
\end{itemize}

Methyl cyanide is a powerful diagnostic probe of the gas temperature in warm dense regions of star formation. \hbox{Millimetre} and submillimetre interferometers such as the Plateau de Bure Interferometer, the Submillimeter Array, the Combined Array for Research in Millimeter-wave Astronomy, and the Very Large Array are capable of resolving the clumpy structure within Orion KL and can provide detailed information on its properties, while the Atacama Large Millimeter Array is now able to observe CH$_3$CN with sufficient sensitivity to discern the morphology of the weaker extended emission. Such follow-up observations will be vital to confirm the shock heating scenario we propose here. Combined with legacy datasets such as those presented in this paper, future observations of methyl cyanide promise to unveil the detailed temperature structure in the very hearts of these star-forming regions.


\begin{acknowledgements}
We are grateful to D.~Lis for providing us with the 350\,$\mu$m continuum map of Orion. We also thank the referee and the editor, M.~Walmsley, for useful comments which helped to improve an earlier draft of this paper. We thank the Spanish MINECO for funding support from grants CSD2009-00038, AYA2009-07304, and AYA2012-32032. TAB is supported by a JAE-DOC research contract. AP is supported by a JAE-DOC CSIC fellowship co-funded with the European Social Fund under the program `Junta para la Ampliaci\'on de Estudios', by the Spanish MICINN grant AYA2011-30228-C03-02 (co-funded with FEDER funds), and by the AGAUR grant 2009SGR1172 (Catalonia). The National Radio Astronomy Observatory is a facility of the National Science Foundation operated under cooperative agreement by Associated Universities, Inc.
\end{acknowledgements}



\appendix

\section{Rotation and population diagrams}\label{Appendix:Analysis-Details}

\subsection{Rotation diagrams}\label{Appendix:Rotation-Diagrams}

The equation for the rotation diagram analysis derives from the Boltzmann equation,
\begin{equation}\label{Equation:Rotation-Diagram}
\ln \left( \frac{N_\mathrm{up}}{g_\mathrm{up}} \right) = \ln \left( \frac{N_\mathrm{tot}}{Q_\mathrm{rot}} \right) - \frac{E_\mathrm{up}}{kT_\mathrm{rot}},
\end{equation}
where $N_\mathrm{up}$ is the column density (in cm$^{-2}$) of the upper state of a given transition, $g_\mathrm{up}$ is the total degeneracy of that upper state, $N_\mathrm{tot}$ represents the total column density of the molecule, summed over all states, $Q_\mathrm{rot}$ is the dimensionless rotational partition function, $E_\mathrm{up}$ is the energy of the upper state, $T_\mathrm{rot}$ is the rotational excitation temperature, and $k$ is the Boltzmann constant. Under the assumption of optically thin emission, the upper state column density of a transition $i \rightarrow j$ is related to the observed integrated intensity $\int\!T\mathrm{d}V$ (in K\,km\,s$^{-1}$) by the following:
\begin{equation}\label{Equation:Upper-Column}
N_\mathrm{up} = \frac{8 \pi k \nu_{i\!j}^2}{h c^3 A_{i\!j}} \int\!T\,\mathrm{d}V \!\times\! 10^5 \quad (\mathrm{cm^{-2}})\,,
\end{equation}
with $\nu_{i\!j}$ being the frequency of the transition (in Hz), $h$ the Planck constant, $c$ the speed of light, $A_{i\!j}$ the Einstein coefficient for spontaneous emission, and $g_\mathrm{up}$ the upper state degeneracy.

By deriving upper state column densities from the observed integrated line intensities using equation~\ref{Equation:Upper-Column} and plotting these against their respective upper state energies, the rotational excitation temperature $T_\mathrm{rot}$ can be inferred from the reciprocal of the slope of a least-square linear fit to the data points. This in turn allows $Q_\mathrm{rot}$ to be calculated (since it is a function of $T_\mathrm{rot}$) and the total molecular column density $N_\mathrm{tot}$ to be derived.

\subsection{Population diagrams}\label{Appendix:Population-Diagrams}

Accounting for the effects of beam dilution and line opacity in the manner described by \citet{Goldsmith1999}, the equation for the upper state column density is modified as follows:
\begin{equation}\label{Equation:Population-Diagram}
\ln \left( \frac{N_\mathrm{up}^\mathrm{obs}}{g_\mathrm{up}} \right) = \ln \left( \frac{N_\mathrm{tot}}{Q_\mathrm{rot}} \right) - \frac{E_\mathrm{up}}{kT_\mathrm{rot}} - \ln \left( \frac{\Omega_\mathrm{beam}}{\Omega_\mathrm{source}} \right) - \ln \left( C_\tau \right),
\end{equation}
where the beam dilution factor for emission from an unresolved source is the ratio of the solid angles subtended by the source and telescope beam, $f_\mathrm{beam}\!=\Omega_\mathrm{source}/\Omega_\mathrm{beam}$, and the opacity correction factor $C_\tau$ is given by $C_\tau = \tau/(1-e^{-\tau})$. The line opacity at the line centre is given by
\begin{equation}\label{Equation:Line-Opacity}
\tau = \frac{c^3 A_{i\!j}}{8 \pi \nu_{i\!j}^3} \ \frac{g_\mathrm{up}\, N_\mathrm{tot}}{\Delta\!V Q_\mathrm{rot}} \ \exp \left( -\frac{E_\mathrm{up}}{kT_\mathrm{rot}} \right) \left[ \exp \left( \frac{h \nu_{i\!j}}{kT_\mathrm{rot}} \right) - 1 \right],
\end{equation}
where $\Delta\!V$ is the observed line width at FWHM.

Since the total column density and rotational temperature both enter into the calculation of the line opacity, they cannot be determined directly in the manner described for rotation diagrams in the previous section. Instead, predicted upper state column densities are calculated for a grid of models covering some parameter space described by rotational temperature, total column density, and beam dilution factor. The predicted upper state column densities are then compared directly to those derived from observations.

Best-fit parameters can then be determined by performing a $\chi^{2}$ minimisation over the entire grid of models, comparing the predicted and observed upper state column densities. We determine the reduced chi-squared for a given set of observed and model values as follows:
\begin{equation}\label{Equation:Chi-Squared}
 \chi^2_\mathrm{red} = \frac{1}{\textit{\#}_\mathrm{obs} - \textit{\#}_\mathrm{par}} \sum_i \left( \frac{N_{\mathrm{up},\,i}^\mathrm{\,obs} - N_{\mathrm{up},\,i}^\mathrm{\,mod}}{\sigma_i^\mathrm{\,obs}} \right)^2.
\end{equation}
Here, $\textit{\#}_\mathrm{obs}$ is the number of observed lines being compared, $\textit{\#}_\mathrm{par}$ is the number of free parameters in the models (four: $N_\mathrm{tot}$, $T_\mathrm{rot}$, $f_\mathrm{beam}$, and $\tau$, with the line width taken directly from the Gaussian fits), $N_{\mathrm{up},\,i}^\mathrm{\,obs}$ and $N_{\mathrm{up},\,i}^\mathrm{\,mod}$ are the observed and model upper state column densities for transition $i$, respectively, and $\sigma_i^\mathrm{\,obs}$ is the 1-sigma uncertainty on the observed upper state column density for that transition (including uncertainties from both the Gaussian fit and observed flux).


\section{Beam dilution maps}\label{Appendix:Beam-Dilution}

\begin{figure*}[!h]
 \centering
 \resizebox{13.5cm}{!}{\includegraphics{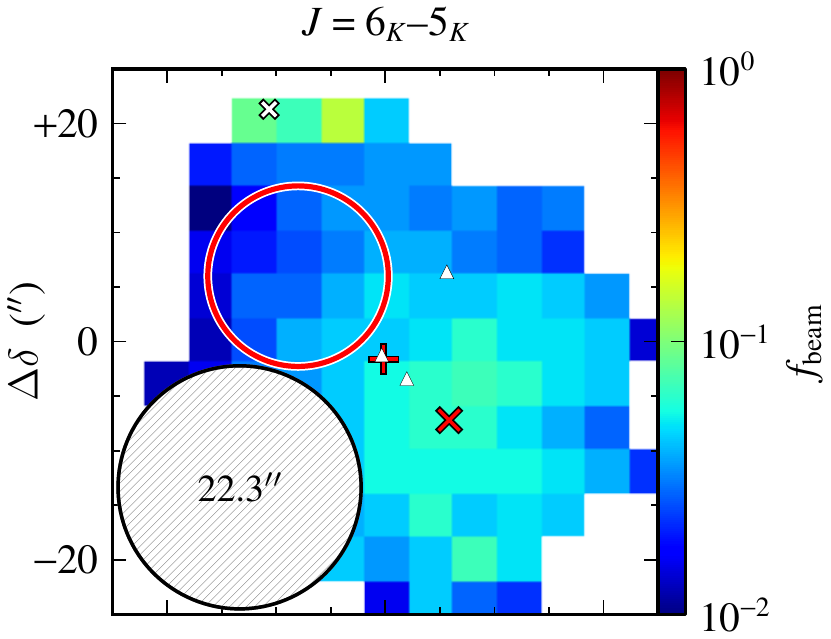}\hspace{5mm}
 \includegraphics{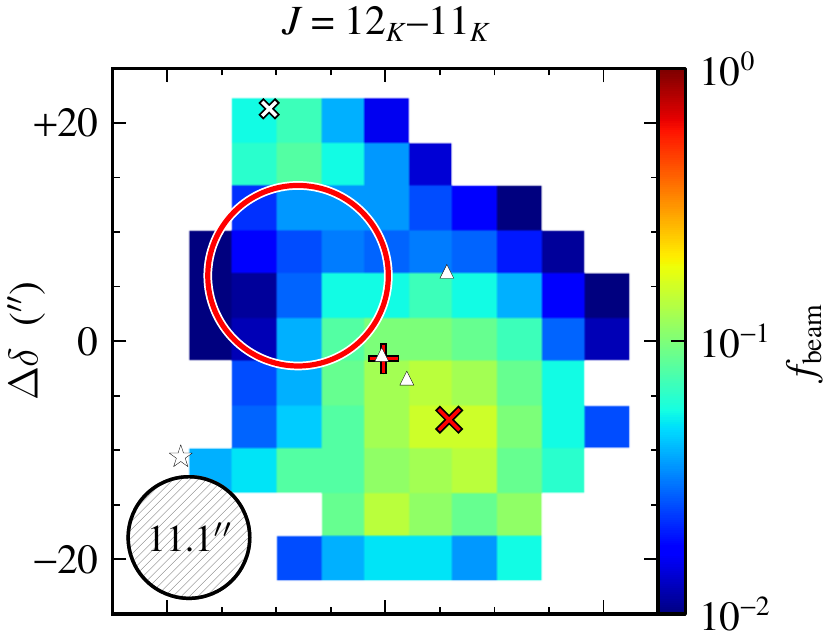}}\\
 \resizebox{13.5cm}{!}{\includegraphics{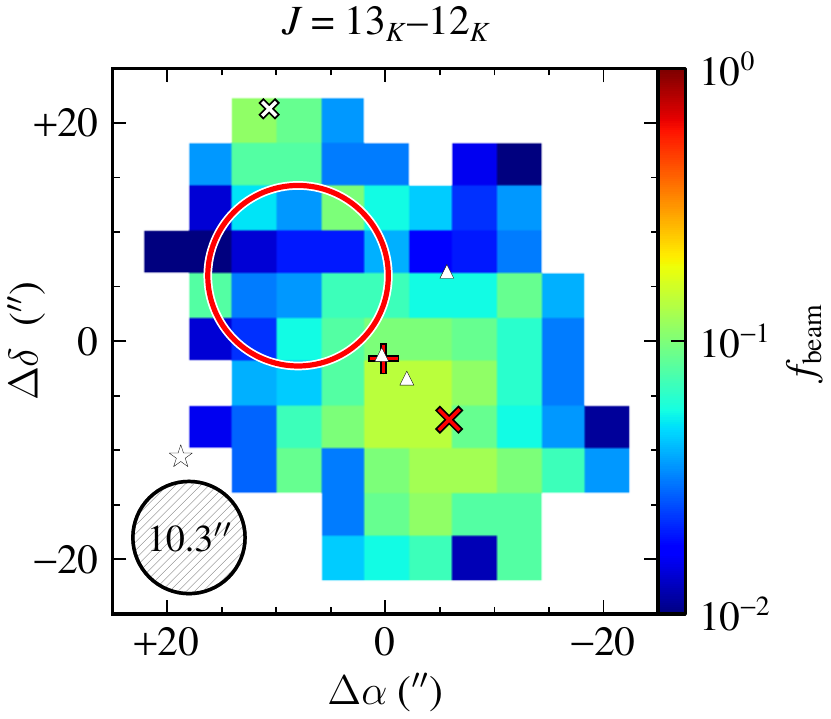}\hspace{5mm}
 \includegraphics{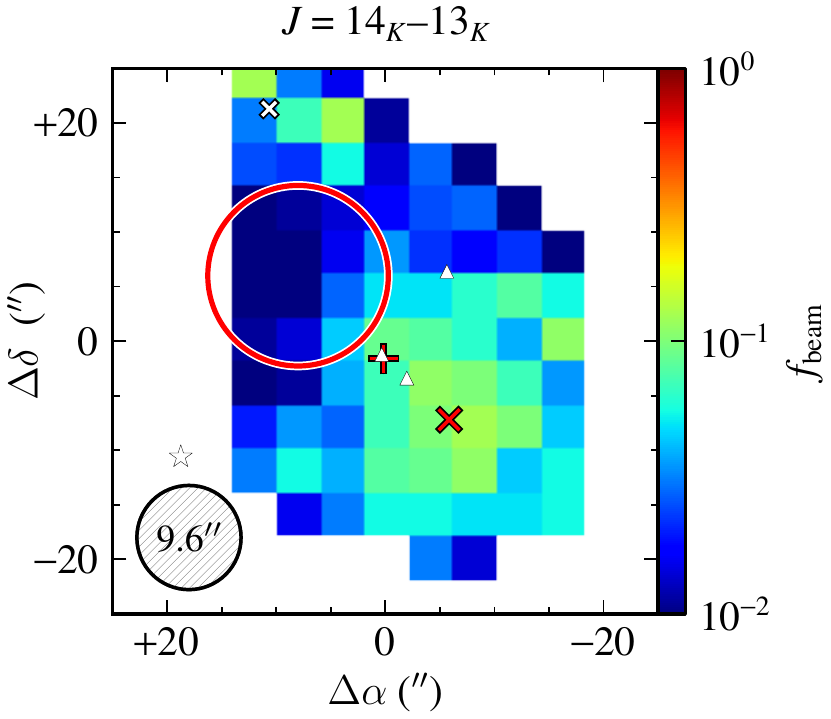}}
 \caption{Maps of the beam dilution factor across the Orion KL region determined by population diagram analysis of the observed lines of CH$_3$CN $J$=$6_K$--$5_K$, $J$=$12_K$--$11_K$, $J$=$13_K$--$12_K$, and $J$=$14_K$--$13_K$ at each map position (see Sect.~\ref{Analysis:Opacity-Dilution} for details). A dilution factor of 0.1 corresponds to a source size of 7$\arcsec$ for the $J$=$6_K$--$5_K$ map and $\sim$3$\arcsec$ for the other maps. The locations of the various sources are indicated as in previous figures.}
 \label{Fig:PD_fbeam_maps}
\end{figure*}
\begin{figure*}
 \centering
 \resizebox{13.5cm}{!}{\includegraphics{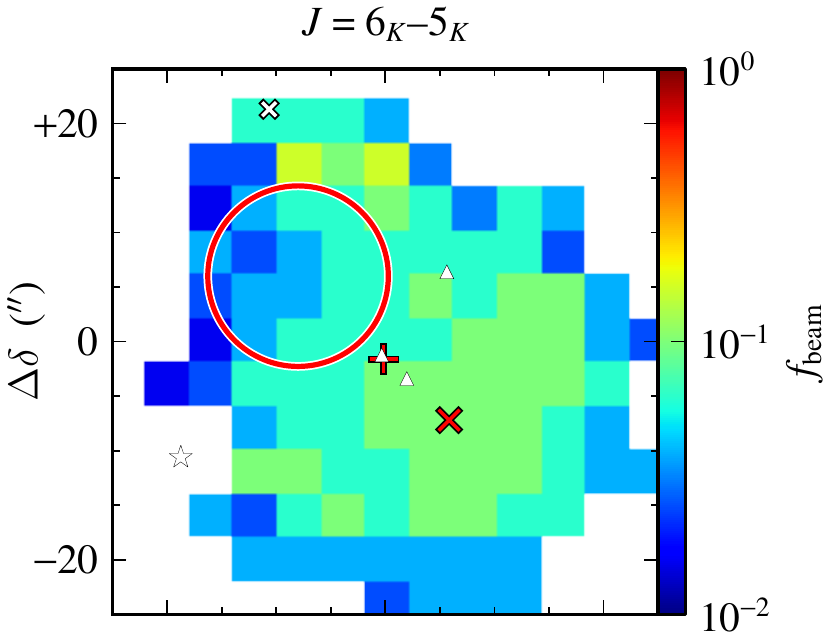}\hspace{5mm}
 \includegraphics{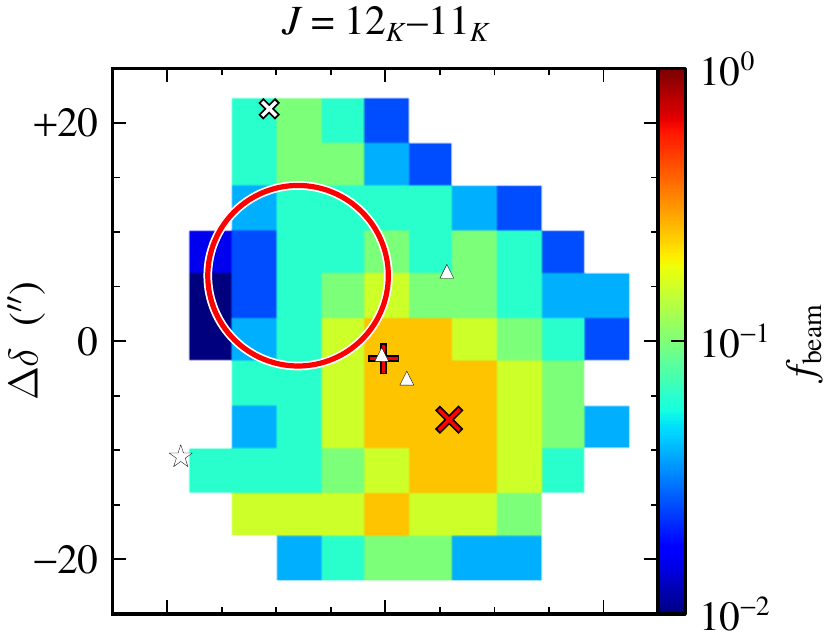}}\\
 \resizebox{13.5cm}{!}{\includegraphics{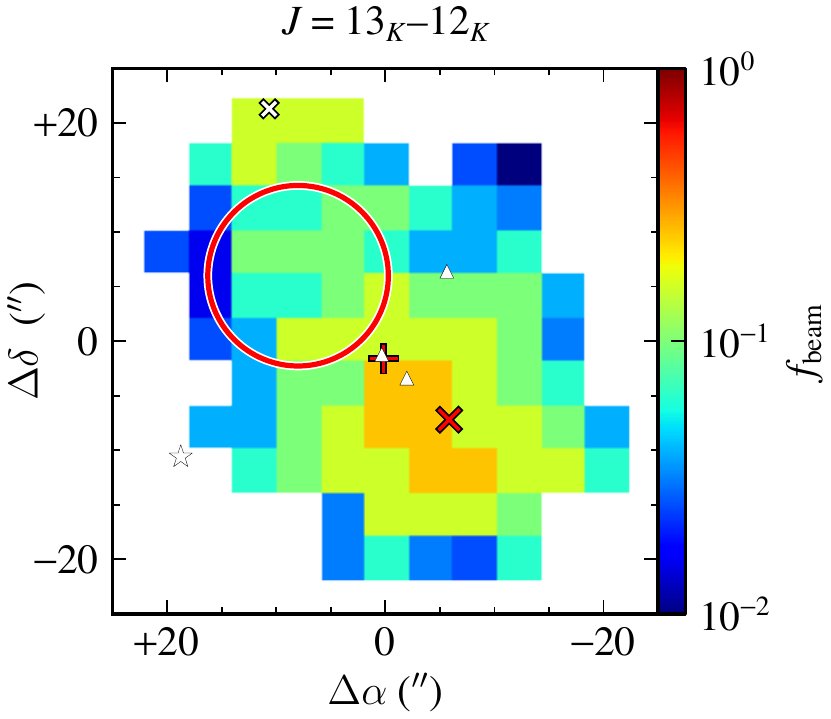}\hspace{5mm}
 \includegraphics{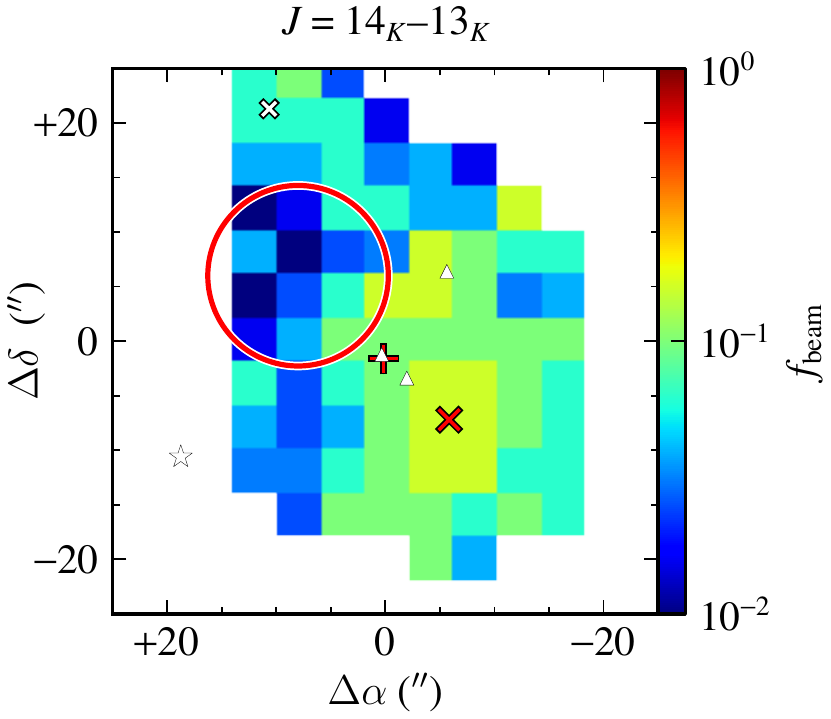}}
 \caption{Maps of the beam dilution factor ($f_\mathrm{beam}$) across the Orion KL region determined by fitting LVG models to the observed lines of CH$_3$CN $J$=$6_K$--$5_K$, $J$=$12_K$--$11_K$, $J$=$13_K$--$12_K$, and $J$=$14_K$--$13_K$ at each map position (see Sect.~\ref{Analysis:LVG-Model-Fits} for details). A dilution factor of 0.1 corresponds to a source size of 7$\arcsec$ for the $J$=$6_K$--$5_K$ map and $\sim$3$\arcsec$ for the other maps. The locations of the various sources are indicated as in previous figures.}
 \label{Fig:LVG_fbeam_maps}
\end{figure*}

\end{document}